\documentclass{aa}
\bibpunct{(}{)}{;}{a}{}{,} 
\usepackage{natbib}
\usepackage{graphicx}
\usepackage{epstopdf}
\usepackage{txfonts}
\usepackage{bm}
\usepackage{amsmath,amssymb}

\usepackage[svgnames]{xcolor}
\definecolor{ku}{RGB}{144,26,30}
\definecolor{ku-yellow}{RGB}{255,249,25}
\definecolor{olive}{RGB}{107,142,35}
\definecolor{olive-yellow}{RGB}{154,205,50}

\begin{document}

\title{Physical properties of terrestrial planets and water delivery in the habitable zone using N-body simulations with fragmentation}
   \author{A. Dugaro \inst{1,2}
           \thanks{adugaro@fcaglp.unlp.edu.ar},
           G. C. de El\'ia \inst{1,2}, \and
           L. A. Darriba \inst{1,2}}

   \offprints{A. Dugaro
    }

  \institute{Instituto de Astrof\'{\i}sica de La Plata, CCT La Plata-CONICET-UNLP \\
   Paseo del Bosque S/N (1900), La Plata, Argentina
   \and Facultad de Ciencias Astron\'omicas y Geof\'\i sicas, Universidad Nacional de La Plata \\
   Paseo del Bosque S/N (1900), La Plata, Argentina}
   
   \date{Received / Accepted}

  \abstract
   {}
   {The goal of this research is to study how the fragmentation of planetary embryos can affect the physical and dynamical properties of terrestrial planets around solar-type stars. Our study focuses on the formation and evolution of planets and water delivery in the habitable zone (HZ). We distinguish class A and class B HZ planets, which have an accretion seed initially located inside and beyond the snow line, respectively. 
   }
   {We develop an N-body integrator that incorporates fragmentation and hit-and-run collisions, which is called D3 N-body code. From this, we perform 46 numerical simulations of planetary accretion in systems that host two gaseous giants like Jupiter and Saturn. We compare two sets of 23 N-body simulations, one of which includes a realistic collisional treatment and the other one models all impacts as perfect mergers.
   }
   {The final masses of the HZ planets formed in runs with fragmentation are about 15 \% - 20 \% smaller than those obtained without fragmentation. As for the class A HZ planets, those formed in simulations without fragmentation experience very significant increases in mass respect to their initial values, while the growth of those produced in runs with fragmentation is less relevant. We remark that the fragments play a secondary role in the masses of the class A HZ planets, providing less than 30 \% of their final values. In runs without fragmentation, the final fraction of water of the class A HZ planets keeps the initial value since they do not accrete water-rich embryos. In runs with fragmentation, the final fraction of water of such planets strongly depends on the model used to distribute the water after each collision. The class B HZ planets do not show significant differences concerning their final water contents in runs with and without fragmentation. From this, we find that the collisional fragmentation is not a barrier to the survival of water worlds in the HZ.
   }
   {}

\keywords{Terrestrial planets - Methods: numerical - Protoplanetary disks}

\authorrunning{A. Dugaro,
           G. C. de El\'ia \and
           L. A. Darriba
               }
\titlerunning{Planets in N-body simulations with fragmentation}

\maketitle
\section{Introduction}

Understanding terrestrial planet formation is an ongoing challenge in planetary sciences. We know that planets are formed around stars as a natural by-product of the star
formation process. We can broadly summarize it in different stages \citep{Morbidelli2012}. Once the star collapses, a disk of gas and dust forms. Via an accretion process in the disk, the dust settles into orbit about the host star. Then, as small grains or ices orbit, they collide at low relative velocity and stick forming aggregates \citep{Weiden1977,Krijt2015}.

While the entire mechanism is not yet fully understood, these aggregates will eventually become bodies of a few kilometers across, called planetesimals. Once they reach this size, the planetesimals are decoupled from the gas and move on Keplerian orbits around the star. Mutual gravitational interactions and collisions then become important as their cross-sections for accretion becomes larger than their physical size as their gravitational pull is sufficient to capture adjacent material. As long as the impacting velocities are smaller than the escape velocity of the larger planetesimals, they can grow rapidly.
When this happens, the planetesimals grow much faster than smaller bodies and lead to a new phase of accretion called the runaway accretion \citep{Kokubo1996}, resulting in planetary embryos or protoplanets of 50+ km in size. Runaway growth breaks the uniformity of the velocity and spatial distributions of planetesimals \citep{Kokubo1998}. The growth of larger embryos slows down when protoplanets grow large enough to increase the velocity dispersion of the planetesimals nearby, reducing the rate of accretion \citep{Ida1993}. Under this scenario, several protoplanets are formed and dominate the planetesimals dynamical evolution. This stage is called oligarchic growth \citep{Kokubo1998}. Finally, the last stage of planetary formation consists of an increase of close encounters that lead to giant impacts between protoplanets. The final configuration of the planetary system is defined at the end of this stage between 10 Myr - 150 Myr \citep{ChamWeth1998,Chambers2001,Jacobson2014}.    

The N-body integrators are known for study the different stages of planetary formation. In particular, a great variety of this kind of integrators exists in the scientific community for different physics problems.
The main goal of performing N-body simulations is to study the late stages of terrestrial planet formation.
Several authors used this kind of integration to analyze the late stage of accretion for the solar system (e.g., \citet{Chambers2001,Obrien2006,Raymond2009}. to name a few) and for exoplanetary systems with different stellar types, and with or without gas giants (e.g., \citet{deElia2013,Dugaro2016,Darriba2017,Zain2018,Sanchez2018}, among others). In these studies, although the dynamical evolution was well represented, the collision treatment was simplified. Colliding planetary embryos perfectly merge into a new body conserving mass and water content. One of the reasons of this simple model is that the computation time in an N-body simulation scales with N$^2$, where N is the number of particles in the system. We have to set a compromise between the number of bodies and performance. Without allowing fragmentation between planetary embryos, the number of bodies in the system will not to be increased. As collisions take place, the number of bodies drops and the computing time decreases.

\citet{Leinhardt2012}, hereafter LS12, presented a complete analytic collision model for gravity-domained bodies. The range of outcomes that LS12 propose includes partial accretion, partial erosion, catastrophic collisions, and hit-and-run encounters.
These outcomes for each collision depend on the target size, projectile size, impact velocity, and impact angle. LS12 define the catastrophic disruption criteria Q$_\text{RD}^*$ as the specific energy needed to disperse half of the involved mass in the collision. For this reason, the authors introduce two material parameters, $\bar{\mu}$ as a coupling parameter and the dissipation parameter $c^*$.
With the analytic derivation and scaling laws presented in their work, LS12 describe the transition between collision regimes and size-velocity distribution of the post-collision bodies.
The analytic model developed by LS12 is a powerful tool that has two major advantages: it improves the physics of collisions in numerical simulations of planet formation and collisional evolution, and is easy to adapt to an N-body code.

On the other hand, \citet{Genda2012} studied the merging criteria for collisions of rocky embryos on terrestrial planet formation. The authors used a smoothed particle hydrodynamic method for the giant impacts and investigated the critical impact velocity, where transition between hit-and-run and merging takes place. The authors derived a simple formula for the normalized critical impact velocity $v_\text{cr}$/$v'_\text{esc}$ (where $v'_\text{esc}$ is the two-body escape velocity), which depends on the mass ratio of the protoplanets and the impact angle.
That said, a hit-and-run collision with an impact velocity lower than the critical value may result in a second collision leading to a perfect merge.

This improvement was adopted by several authors to study the solar and extrasolar systems. \citet{Chambers2013} implemented a refined collisional algorithm into MERCURY based on the work developed by LS12. In that work he studied the final stage of terrestrial planet formation with N-body simulations, concluding that hit-and-run collisions are a common outcome of two colliding big bodies. The author came to this conclusion comparing the dynamical evolution of two identical sets of initial conditions, both considering and not considering fragmentation. He found two important differences between these two models. On the one hand, the planetary masses are lower when fragmentation is included. On the other hand, the excitation of the eccentricity over time is also lower in comparison with the model that considers only perfect merges.

\citet{Quintana2016} performed N-body simulations to study the terrestrial planet formation around a sun-like star, using the modified version of MERCURY implemented on \citet{Chambers2013}. The authors compared the standard accretion model with the modified version of MERCURY \citep{Chambers2013}, that allows fragmentation in the collisions between big bodies.
The authors have found that the collisional history of the planets that survive at the end of the integration, differs significantly between the two models, although the overall masses and number of planets generated are comparable.

More recently, \citet{Mustill2018} carried out his own implementatation of the collisional algorithm developed by LS12 into the public version of MERCURY. In particular, these authors studied how this improvement in the treatment of the collisions affects the outcomes concerning the in-situ planet formation on packed systems and instability scenarios. It is worth noting that, unlike the works carried out by \citet{Chambers2013}, \citet{Quintana2016}, and \citet{Wallace2017}, \citet{Mustill2018} introduced a factor of mass removal in the collisional algorithm in order to represent the material that is ground and then removed from the system by radiation forces.

With this in mind, it is evident the need to have a numerical tool of our own that allows us to study in a more detailed way the collisional history of the bodies that make up a planetary system as well as its dynamical evolution. Thanks to the improvements in the collisional model proposed by LS12, we decided to move away from the perfect merging model and developed the D3 N-body code with a more realistic treatment of collisions between planetary embryos. This improvement allows us to carry out a detailed study about the final composition of the planets formed. In particular, we can study the water delivery in a more realistic way than in the classic models of accretion.

In this sense, \citet{Marcus2010} described two empirical models for the mantle stripping in differentiated planetary embryos after a collision. The authors used a simple differentiated structure for the terrestrial embryos, assuming an iron core and an ice/silicate mantle, and performed a series of SPH simulations with the GADGET code \citep{Springel2005}. In particular, \citet{Marcus2010} varied the projectile-to-target ratio, the impact angle, and the number of particles of each body in order to investigate how the giant impacts affect the final abundance of water in such bodies. From this, they concluded that the more violent the collisions, the more mantle is lost, and that water is more easily removed than the silicates. 

More recently, \citet{Dvorak2015} obtained interesting results concerning to water content retained in significant fragments after a collision. From SPH simulations, they studied the outcome of an impact between two bodies with different values of the impact velocity and the impact angle. One relevant conclusion of their research suggests a significant water loss for faster and/or less inclined collisions. 

The investigations developed by \citet{Marcus2010} and \citet{Dvorak2015} suggest that to incorporate a realistic model of transport and removal of volatiles in an N-body code may lead to reduced water contents of the resulting terrestrial-like planets in comparison with those derived from classical models that assume perfect mergers.  

The main goal of the present research is to study the physical and dynamical properties of terrestrial-like planets and water delivery in the habitable zone (HZ) using N-body simulations that incorporate fragmentation and hit-and-run collisions. According to this, this paper is therefore structured as follows. In Sect.~\ref{sec:nbody-code}, we present the main properties of the D3 N-body code integrator and the collisional model introduced by LS12. In Sect.~\ref{section:applications}, we describe the application and the initial conditions used to carry out our simulations. In Sect.~\ref{section:results}, we show our results and carry out a detailed analysis of all simulations and the planets remaining in the HZ. Finally, we discuss such results within the framework of current knowledge of planetary systems and the limitations of the model in Sect.~\ref{section:discusion}.

\section{N-body Code: Classic integrators}
\label{sec:nbody-code}
In this section, we describe some basic ideas of the symplectic theory and the general properties of the numerical integrator.
\subsection{Overview}
The solution to the N-body problem is obtained through the resolution of the equations of motion that derive from the system's Hamiltonian.
In effect, the Hamiltonian of a system of N + 1 particles of masses $m_i$ ($i = 0,...,N$), subject solely to the action of their mutual gravitational attraction, is given by

\begin{equation}\label{eq:1}
H(\bm{q},\bm{p}) = \sum\limits_{i=0}^N \frac{| \bm{p}_{i}|^{2}}{2 m_{i}} - G \sum\limits_{i=0}^{N-1} \sum\limits_{j=i+1}^N \frac{m_{i} m_{j}}{|\bm{q}_{i}- \bm{q}_{j}|},
\end{equation}

\noindent{where} $\bm{q}_{i} = (x_\text{i},y_\text{i},z_\text{i})$ and $\bm{p}_\text{i} = (p_{x_\text{i}},p_{y_\text{i}},p_{z_\text{i}})$ are the position and momentum vectors of the $i$th body (of mass $m_i$) with respect to an inertial reference frame, and $G$ the universal gravitational constant.

However, for the particular problem of a planetary system, where most of the mass is concentrated in one body (e.g. the sun in the solar system), it is more convenient to divide the Hamiltonian as follows:
\begin{equation}
  H = H_\text{Kep} + H_\text{int},
  \end{equation}

\noindent{where} $H_\text{Kep}$ is the part of the Hamiltonian that describes the Keplerian motion of the particles around the massive body and $H_\text{int}$ is the part that describes the interaction between each pair of particles, except with the main body (i.e. the star).

Depending on the coordinates chosen, the Hamiltonian can be divided in different ways.
\citet{Wisdom1991}, for instance, used Jacobian coordinates. This system constitutes a reference frame where the position and momentum of each body are considered with respect to the center of mass of all the bodies with indices lower than a certain previously defined value. Symplectic integrators based on this division of the Hamiltonian are known as symplectic integrators of mixed variables (mixed variable symplectic methods) \citep{Saha1992}.

An alternative procedure proposed by \citet{Duncan1998} is to define the position with respect to the main body and the momentum with respect to the barycenter of the system. This set of variables is defined as heliocentric democratic variables. By means of a canonical transformation, it is possible to transform the original set of coordinates and momenta ($\bm{q}$, $\bm{p}$) into a new set of heliocentric coordinates ($\bm{Q}$) and barycentric momenta ($\bm{P}$).

Implementing this transformation, and adopting this new set of variables, the Hamiltonian from Eq.~\eqref{eq:1} could be rewritten in the following way

\begin{align}
  H(\bm{Q},\bm{P}) &=  \sum\limits_{i=1}^{N} \left( \frac{|\bm{P}_{i}|^{2}}{2m_{i}} - \frac{G m_{i} m_{0}}{|\bm{Q}_{i}|}\right) - G \sum\limits_{i=1}^{N-1} \sum\limits_{j=i+1}^N \frac{m_{i} m_{j}}{|\bm{Q}_{i}-\bm{Q}_{j}|} \nonumber \\
  &+ \frac{1}{2m_{0}} \left|\sum\limits_{i=1}^{N} \bm{P}_{i}\right|^{2} + \frac{|\bm{P}_{0}|^{2}}{2 M}.\label{eq:HQP}
\end{align}

Since in Eq. \eqref{eq:HQP} the coordinates $\bm{Q}_i$ are heliocentric, that corresponding to the central body vanishes, i.e: $\bm{Q}_0 = 0$. Consequently, the Hamiltonian is independent of the coordinate $\bm{Q}_0$, therefore the momentum $\bm{P}_{0}$ is a constant of motion, and its contribution to the Hamiltonian is not considered. It is worth noting that $N$ indicates the number of particles excluding the star and $m_0$ indicates its mass. Therefore, we can group the terms from Eq.~\eqref{eq:HQP} as

\begin{equation}
H(\bm{Q},\bm{P}) = H_{\text{Kep}} + H_{\text{int}} + H_{\odot},
\end{equation}

\vspace*{0.5cm}

\noindent{where}

\begin{align}
H_{\text{Kep}} &= \sum\limits_{i=1}^{N} \left( \frac{|\bm{P}_{i}|^{2}}{2m_{i}} - \frac{G m_{i} m_{0}}{|\bm{Q}_{i}|}\right), \label{eq:Hkep} \\
H_{\text{int}} &= - G \sum\limits_{i=1}^{N-1} \sum\limits_{j=i+1}^N \frac{m_{i} m_{j}}{|\bm{Q}_{i}-\bm{Q}_{j}|}, \label{eq:Hint}\\
H_{\odot}     &= \frac{1}{2m_{0}} \left|\sum\limits_{i=1}^{N} \bm{P}_{i}\right|^{2}\label{eq:Hsun},
\end{align}

\noindent{each} one of these terms describing the following:

\begin{itemize}

\item[$\star$] $H_\text{Kep}$: the Keplerian motion of the particles around the massive body,
  \vspace{3mm}
\item[$\star$] $H_\text{int}$: the interaction between each pair of particles (except with the main body),
  \vspace{3mm} 
\item[$\star$] $H_{\odot}$: the main body barycentric momentum, which appears due to the chosen set of variables. 
\end{itemize}

Following \citet{Chambers1999}, by solving Hamilton's equations of motion, the general solution to the rate of change of a given function, $u = u(q,p)$, over a time $\tau$, starting from its initial value $u(0)$, is given by:
\begin{equation}
  u(\tau) = e^{\tau H}u(0),
  \label{eq:u_tau}
\end{equation}

\noindent
where $H$ is the Hamiltonian operator. As shown in the aforementioned paper, if an operator $H$ is written as the sum of two operators ( i.e: $H = H_\textrm{A} + H_\textrm{B}$), then that in Eq. \eqref{eq:u_tau} can be split as the consecutive product of the two operators,
\begin{equation}
  u(\tau) = e^{\tau (H_\textrm{A} + H_\textrm{B})} u(0) = e^{\tau H_\textrm{A}} e^{\tau H_\textrm{B}} u(0).
\end{equation}

By splitting $H_\textrm{int}$ and $H_\odot$ in half, we obtain a second-order integrator, and Eq. \eqref{eq:u_tau} becomes

\begin{equation}\label{eq:symplectic_operator}
  u(\tau) = e^{\tau H_\text{int}/2} e^{\tau H_{\odot}/2} e^{\tau H_\text{Kep}} e^{\tau H_{\odot}/2} e^{\tau H_\text{int}/2} u(0).
\end{equation}

From the operator defined in Eq.~\eqref{eq:symplectic_operator}, we can construct a 5-step integration scheme as follows:

\begin{enumerate}
 \item[I)] the coordinates remain fixed and each body receives an acceleration from the other bodies (but not from the main body) that modifies its momentum over a time interval $\tau/2$;
 \item[II)] the momenta remain fixed, and each body undergoes a displacement in its position in the amount $(\tau/2m{_0}) \sum\limits_{i=1}^{N}\bm{P}_{i};$
 \item[III)] each body evolves around a Keplerian orbit (with the same central location and the same central mass) over the whole time interval $\tau$;
 \item[IV)] as step II;
 \item[V)] as step I.
\end{enumerate}
  
However, if two bodies are too close to each other, the corresponding term in $H_\text{int}$ becomes large enough so $H_\text{Kep}$ ceases to be the dominant term and the error increases substantially. For this reason, the previous scheme can not resolve situations of close encounters.

One of the solutions proposed to address this problem \citep{Duncan1998} consists in dividing the perturbative terms of $H_\text{int}$ and assigning each of them a different integration step-size, so that the strongest perturbations have the smallest step-sizes. The resulting integrator is purely symplectic, although complicated to carry out, and also does not retain the high speed of the basic symplectic method.
An alternative solution is proposed by \citet{Chambers1999}, who constructs a hybrid algorithm that mixes both symplectic and non-symplectic components in order to retain properties of both. As we mentioned before, we need to make the term $H_\text{int}$ small again (so that $H_\text{Kep}$ remains dominant). One way to do this is to transfer the close encounter term in $H_\text{int}$ to $H_\text{Kep}$ for the duration of the close encounter. For example, if the bodies $\alpha$ and $\beta$ have a close encounter, the terms from Eq. \eqref{eq:Hkep} and \eqref{eq:Hint} result in 
\begin{align}
  H_{\text{Kep}} &= \sum\limits_{i=1}^{N} \left( \frac{|\bm{P}_{i}|^{2}}{2m_{i}} - \frac{G m_{i} m_{0}}{|\bm{Q}_{i}|}\right) - \frac{G m_{\alpha} m_{\beta}}{|\bm{Q}_{\alpha} - \bm{Q}_{\beta}|},\\
H_\text{int}  &= - G \sum\limits_{i=1}^{N-1} \sum\limits_{j=i+1}^N \frac{m_{i} m_{j}}{|\bm{Q}_{i}-\bm{Q}_{j}|} -  \sum\limits_{\substack{j=\alpha+1 \\ j\neq \beta}}^{N-1}\frac{G m_{\alpha} m_{j}}{|\bm{Q}_{\alpha} - \bm{Q}_{j}|}.
\end{align}

With this scheme, $H_{\text{Kep}}$ can no longer be solved analytically. However, we can solve it numerically using a conventional integrator (briefly described later), and the remaining bodies that do not have close encounters can be analytically integrated.
We need to keep in mind that doing this brings a modification in the original Hamiltonian and the integrator no longer remains purely symplectic. In order to make the hybrid integrator symplectic, we have to make sure that no term is transferred between the different parts of the Hamiltonian. Following \citet{Chambers1999}, we can do this by dividing each term of interaction between $H_\text{Kep}$ and $H_\text{int}$ so that the corresponding part in $H_\text{int}$ is always kept small, while the part in $H_\text{Kep}$ is evaluated only during close encounter:

\begin{align}
  H_{\text{Kep}} &= \sum\limits_{i=1}^{N} \left( \frac{|\bm{P}_{i}|^{2}}{2m_{i}} - \frac{G m_{i} m_{0}}{|\bm{Q}_{i}|}\right) \nonumber \\ &- G \sum\limits_{i=1}^{N-1} \sum\limits_{j=i+1}^N \frac{m_{i} m_{j}}{|\bm{Q}_{i}-\bm{Q}_{j}|}(1 - K(|\bm{Q}_{i}-\bm{Q}_{j}|)),\\
H_{\text{int}}  &= - G \sum\limits_{i=1}^{N-1} \sum\limits_{j=i+1}^N \frac{m_{i} m_{j}}{|\bm{Q}_{i}-\bm{Q}_{j}|} K(|\bm{Q}_{i}-\bm{Q}_{j}|).
\end{align}

The form of the function $K$ must be such that, when the separation $|\bm{Q}_{i}-\bm{Q}_{j}|$ between the bodies $i$ and $j$ is large, $K$ must tend to one, while it tends to zero when such a separation is small. On the other hand, the transition has to be smooth enough. Following \citet{Chambers1999}, we adopt the function,
\begin{displaymath}
K = 
  \begin{cases}
0 \ \ \qquad \qquad \qquad \ \ \text{if}  \ \ y < 0 \\
y^2/(2y^2-2y+1) \ \ \text{if} \ \  0 < y < 1\\
1 \ \ \qquad \qquad \qquad \ \ \text{if} \ \ y > 1
\end{cases}
\end{displaymath}

\noindent{where}

\begin{equation}
    y = \left( \frac{r_\text{ij} - 0.1 r_\text{crit}}{0.9 r_\text{rcrit}} \right)
\end{equation}

and $r_\text{crit}$ it is a free parameter that indicates the critical switching distance. Generally is a multiple of the mutual Hill radius of the involved bodies.

This ensures that that $|H_\text{int}| \le |H_\text{Kep}|$, even during a close encounter (for a detailed explanation, refer to \citet{Chambers1999}).
Nevertheless, the choice of $K$ is an open debate on how the integrator keeps its symplecticity \citep{HernandezA2019, Rein2019} about the switchover function and its expression is discussed in Sect.~\ref{section:discusion}.
Thus, the second-order hybrid integrator proceeds as follows:

\begin{enumerate}
\item[I)] the coordinates remain fixed and each body receives an acceleration from the other ones (but not from the main body), weighted in the situations of close encounter by a factor $K$, which modifies its momentum over a time interval $\tau/2$;
\item[II)] the momenta remain fixed, and each body undergoes a displacement in its position in the amount $(\tau/2m{_0}) \sum\limits_{i=1}^{N}\bm{P}_{i}$ over all the $N$ bodies (except the central star);
\item[III)] the bodies that are not in close encounter move in a Keplerian orbit around the main body over a time interval $\tau$. For those that are in close encounter, the Keplerian terms and the close encounter terms are weighted by (1 - $K$) and integrated numerically over a time interval $\tau$;
\item[IV)] as step II;
\item[V)] as step I.
\end{enumerate}

It is worth noting that the accretion problems that we focus in this work take place in step III, where the collision conditions are given as a result of the numerical integration.

Following \citet{Chambers1999}, we adopt the same numerical integrator for solving the close encounters. Because of its robustness, speed and precision, the Bulirsch-St\"oer method is an excellent choice for this kind of numerical problems. 

\subsection{Collisions in classic integrators}

Perfect merging is the simplest collision model used in simulations of planetary accretion studies so far, either to study the solar system \citep{Chambers2001, Obrien2006, Raymond2009} as well as extrasolar systems \citep{deElia2013, Dugaro2016, Darriba2017, Sanchez2018, Zain2018}. This collisional model does not have physical and geometric parameters involved in the outcome of the collision: two bodies that collide will result in a body with a mass equal to the sum of the masses of the target and the projectile.

Recently, LS12 performed high-resolution simulations of collisions between planetesimals. There, they found a wide range of possible outcomes: merging, cratering, super-catastrophic disruption, and hit-and-run events. The authors derived useful scaling laws that determined the transition between the different regimes. The different collision outcomes were determined by physical properties such as impact velocity, mass ratio of the bodies, and a geometric parameter like the impact angle.

It is important to remark that, in the D3 N-body code, embryos are treated as big bodies, meaning that they interact with any other body in the system. Planetesimals and fragments, on the contrary, are treated as small bodies, that is, they don't interact with each other, only feeling the gravitational pull from the embryos and the central star.

\subsection{The collisional model - Different type of collisions}
\label{section:collisional_model}

In this section we briefly describe the regimes derived by LS12 and were used in this work. After this description, we present the algorithm to identify and calculate the different giant impacts. 

Collision outcomes can be described in terms of the impact energy per unit mass $Q$ and a critical value $Q_\text{RD}^{*}$, where $Q_\text{RD}^{*}$ is defined as the specific energy per unit mass required to disperse half of the total colliding mass. Depending on the collision, it may result in a large body and several fragments. In this case, we will reference as $M_\text{lr}$ to the mass of the largest remnant of the mentioned collision, and the remaining mass is distributed in equal-mass fragments. According to \citet{Chambers2013}, the number of fragments generated will be directly related to a parameter that indicates the minimum permitted fragment mass, $M_\text{min}$. On the one hand, this value has to be large enough so the number of fragments generated is not too high and, thus, slows down the simulations. On the other hand, if the value is too large, the fragments generated will be unrealistic. Therefore, the value $M_\text{min}$ must be set considering a compromise between number of fragments generated, performance and realism of the model.

\begin{figure}
 \centering
 \includegraphics[angle=0, width= 0.48\textwidth]{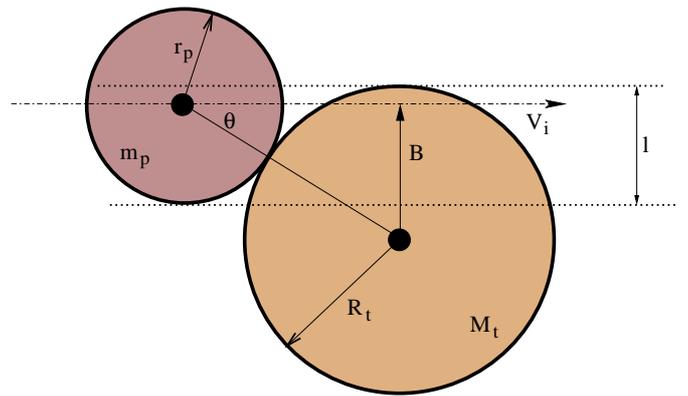}
 \caption{Schematic representation of a collision between two planetary bodies. The target, with a mass $M_\text{t}$ and a radius $R_\text{t}$ collides with a projectile of mass $m_\text{p}$ and radius $r_\text{p}$ at an impact velocity $v_i$. The angle $\theta$ represents the angle formed between $v_i$ and the line connecting the two centers.
}
\label{fig:esquema_colision}
\end{figure}

Let's consider a collision between two planetary bodies as shown in Fig. \ref{fig:esquema_colision}. A target of mass $M_\text{t}$ with a radius $R_\text{t}$ and a projectile of mass $m_\text{p}$ with a radius $r_\text{p}$, where  $M_\text{t}$ $>$ $m_\text{p}$, collide with an impact velocity $v_\text{i}$ and an impact angle $\theta$, where $\theta$ is the angle formed between the line connecting the centers of the two bodies and the projectile velocity vector relative to the target. The bulk density for both the target and the projectile are $\rho_\text{t}$ and $\rho_\text{p}$, respectively. We can describe the different types of collisions as follows:

\begin{itemize}

\item\textbf{Perfect merging}

This type of collision is the most commonly used by the classic integrators so far. No fragments are generated and the remaining body contains the total colliding mass.
\vspace{0.5mm}

\item\textbf{Partial accretion}

When $M_\text{lr}$ is calculated and is larger than the mass of the target but lower than the total colliding mass (i.e: $M_\text{t} \leq$ $M_\text{lr} \leq$ $M_\text{t}$ + $M_\text{p}$), the outcome is considered a partial accretion. The remaining mass, if it's larger than the minimum fragment mass $M_\textrm{min}$, is distributed in fragments. Otherwise, the collision is considered as a perfect merging.

\vspace{0.5mm}

\item\textbf{Erosive collision}

In this case, $M_\text{lr}$ is lower than the mass of the target. The projectile is completely destroyed and, in addition to the remaining mass of the target, is distributed in fragments.

\vspace{0.5mm}

\item\textbf{Super-catastrophic collision}

  In giant impacts, where the impact energy is so large that the
  $M_\text{lr}$ is less than 10\% of the total colliding mass, the outcome is considered a super-catastrophic collision, and the remaining mass is distributed in fragments.

\vspace{0.5mm}

\item\textbf{Hit-and-run}

This outcome occurs when both the target and the projectile remain intact. No fragments are generated.

\vspace{0.5mm}

\item\textbf{Graze and merge}

Depending on the velocity of a hit-and-run collision, it may result in a second collision leading to a perfect merge, with no generation of fragments.

\vspace{0.5mm}

\item\textbf{Erosive hit-and-run}

Called this way because, during a hit-and-run collision, the projectile may be partially or totally disrupted and the target comes out unaffected. The mass stripped out of the projectile is distributed in fragments.

\end{itemize}

With the different types of collisions described, in the next section we will address the modification that has to be made to an N-body numerical integrator, in order to implement the collision classification.

\subsection{Modification in the numerical integrator}

In order to calculate the results obtained by LS12 and apply the classification of the different regimes, we must have detailed information of each collision. More precisely, we need the position of the bodies, the impact velocity, and the impact angle at the time of the collision within a certain tolerance.

Following \citet{Chambers1999}, the collisions are detected with a pre-cheker, based on a Hermite interpolation scheme. When this happens, following \citet{Mustill2018}, we redo the integration in an iterative fashion, by successively dividing the integration step by half, until the bodies are in collision or the integration step is lower than 10$^{-5}$ days. In this way, we have information accurate enough to be able to resolve the collision and classify it with the regime scheme of LS12.

\subsection{The collisional algorithm}
\label{section:collisional_algorithm}

The collisional algorithm implemented in this work is based on LS12 (for a detailed explanation, see the appendix in the aforementioned paper). This algorithm flow can be summarized in a series of steps, as detailed below

\begin{enumerate}

\item When a collision is detected, as one of the bodies, namely the target, is necessarily an embryo (since interaction between planetesimals are neglected), the code first inquires about the body type of the second one involved, i.e: the projectile. If it is a planetesimal (or a fragment), the code assumes a perfect merge, preserving the total mass and momentum. Then, it proceeds to analyze the next collision, starting again on step 1. If the projectile is an embryo, it proceeds to the next step.

\item If, in effect, the collision occurs between two embryos, the mutual escape velocity is calculated by means of the following expression
  \begin{equation}
    v_\text{esc} = \sqrt{\frac{2G(M_\text{t} + m_\text{p})}{R_\text{e}}},
  \end{equation}
  \noindent{where} $R_\text{e} = 3(M_\text{t}/\rho_{t} + m_\text{p}/\rho_{p})/(4\pi)^{1/3}.$ We change the definition of the mutual escape velocity proposed by LS12 and instead we use the one proposed by \citet{Mustill2018}.

\item The comparison between the impact velocity $v_\text{i}$ and the escape velocity $v_\text{esc}$ is made. If $v_\text{i} \le  v_\text{esc}$, then the collision is also assumed to be a perfect merge, conserving the total mass and momentum, as in step 1. In the case $v_\text{i} >  v_\text{esc}$, we proceed to the next step.

\item {We calculate the impact parameter $b = \sin \theta$ and the critical value $b_\text{crit} = R_\text{t}/(R_\text{t}+r_\text{p})$. Depending on the relation between $b$ and $b_\text{crit}$ the collision may result in either a grazing or a non-grazing impact. If $b < b_\text{crit}$ we have a non-grazing impact, and we proceed to step 5. Otherwise, a grazing impact occurs and we skip to step 8.}

\vspace{3mm}
  
\item \emph{Non-grazing} impact ($b < b_\text{crit}$):
  
Calculate the impact energy per unit mass, $Q$, and the catastrophic disruption criterion, $Q_\text{RD}^{*}$. The value of $Q$ is given by
  \begin{equation}
    Q = \frac{1}{2}\mu\frac{v_\text{i}^2}{M_\text{t}+m_\text{p}},
  \end{equation}
  \noindent{where} $\mu$ = $M_\text{t}m_\text{p}/(M_\text{t}+m_\text{p})$. Then, $Q_\text{RD}^*$ is calculated as follows
  \begin{equation}
    Q_\text{RD}^* = \left( \frac{\mu}{\mu_{\alpha}}\right)^{2-3\vec{\bar{\mu}}/2} \left(\frac{c^{*}\pi\rho_{1}G}{5\gamma}\right)\left[R_\text{c1}(1 + \gamma)\right]^{2},
  \end{equation}
  \noindent{where}
  \begin{equation}
    \mu_{\alpha} = \frac{\alpha M_\text{t}m_\text{p}}{M_\text{t}+\alpha m_\text{p}},
  \end{equation}
  \noindent{being $\alpha$} the fraction of the projectile that intersects the target in an oblique impact, $\bar{\mu}$ a measure of how energy and momentum from the projectile are coupled to the target, $c^*$ an energy dissipation factor within the target, $\rho_{1}$ = 1 g/cm$^3$, $\gamma = m_\text{p}/M_\text{t}$, and $R_\text{c1}$ the spherical radius of a body with mass $M_{\text{tot}} = M_\text{t}+m_\text{p}$ with density $\rho_{1}$ \citep{Stewart2009}. According to such a work, here, $\bar{\mu}$ and $c^*$ adopt values of 1/3 and 1.8, respectively.

\item{With the values of $Q$ and $Q_\text{RD}^*$ we check, by simple arithmetic comparison, if the collision is in the super-catastrophic regime ($Q > 1.8 Q_\text{RD}$). This result affects on how to compute the mass of the largest remnant. Following LS12, if the collision is in such a regime, we calculate $M_\text{lr}$ as 
  
\begin{equation}
    M_\text{lr} = 0.1 M_\text{tot} \left(\frac{Q}{1.8Q_\text{RD}^*}\right)^{-3/2},
\end{equation}
  
\noindent
and the mass to distribute in fragments is $m_\text{frag} = M_\text{tot} - M_\text{lr}$. 

\item In the case $Q < 1.8Q_\text{RD}^*$, the collision is not in the super-catastrophic regime, and the mass of the largest remnant is computed as

\begin{equation}
    M_\text{lr} = M_\text{tot}\left(1 - \frac{Q}{2Q_\text{RD}^*}\right).
\end{equation}

By comparing the largest remnant's mass $M_\text{lr}$ and the target's mass $M_\text{t}$, the collision may result in:

 \begin{enumerate}
     \item if $M_\text{lr} \leq M_\text{t}$, it's a partial erosion,
     \item if $M_\text{lr} > M_\text{t}$, it can result in either a partial accretion, if the number of fragments is larger than 0, or a perfect accretion, if $m_{\text{frag}} < M_{\text{min}}$, for which no fragments are generated. 
 \end{enumerate}

}

\vspace{3mm}
  
\item \emph{Grazing} impacts:

The collision is checked to be a possible hit-and-run. If it is not classified as a hit-and-run, in order to classify the collision as super-catastrophic, partial erosion, partial accretion of perfect accretion, we follow the same criteria used for non-grazing impacts, described on steps 5 through 7.

\vspace{3mm}

\item In case the collision is classified as a possible hit-and-run, \citet{Genda2012} investigated the critical impact velocity $v_\text{cr}$, that establishes the boundary between the pure hit-and-run and merging impacts. This velocity is given by
\begin{equation}
  v_\text{cr} = v_\text{esc}'\left[ c_1\Gamma\Theta^{c_5} + c_2\Gamma + c_3\Gamma\Theta^{c_5} + c_4\right],
\end{equation}

\noindent{where} $\Gamma = (1 - \gamma)/(1 + \gamma)$ and $\Theta = 1 - \sin \theta$. The fitting parameters are $c_1 = 2.43$, $c_2 = -0.0408$, $c_3 = 1.86$, $c_4 = 1.08$, and $c_5 = 2.5$. They found that the normalized $v_\text{cr}/v_\text{esc}'$ does not depend on $M_\text{t}$ but rather on $\gamma$ and the impact angle $\theta$, where $v_\text{esc}'$ is the mutual escape velocity.

We applied the formula derived by \citet{Genda2012} and classified the hit-and-run at low velocity impacts as merging impacts. \citet{Chambers2013} coined the term for this type of collision as Graze-and-Merge, meaning that a grazing impact at low velocity derives in a secondary collision where the two bodies merge in one, like a two-stages perfect merging.

Then, if the impact velocity satisfies that $v_\text{i} < v_\text{cr}$, we classified the collision as a Graze-and-Merge.

\item If the collision effectively derives in a hit-and-run (i.e: $v_\text{i} > v_\text{cr}$), the mass of the target is modified in a negligible amount, so we can consider $M_\text{lr}$ = $M_\text{t}$ \citep{Asphaug2006,Genda2012} and we have to calculate the critical disruption energy for the reverse impact on the projectile (See Section 4.2 of LS12). Thus, basically, the roles of the target and projectile are inverted. According to LS12, we set the reverse variables as follows:

\begin{align}
    M_\text{t}^\dagger &= m_\text{p}\\
    m_\text{p}^\dagger &= \eta M_\text{t}
\end{align}

\noindent{where} $\eta$ is the fraction of the target that interacts with the projectile (for a detailed explanation on how to get this fraction see Sect. 4.2 of LS12).
\begin{equation}
  \mu^\dagger = \frac{M_\text{t}^\dagger m_\text{p}^\dagger}{M_\text{t}^\dagger+m_\text{p}^\dagger},
\end{equation}
\begin{equation}
  \gamma^\dagger = \frac{m_\text{p}^\dagger}{M_\text{t}^\dagger},
\end{equation}
\begin{equation}
    Q_\text{RD}^\dagger = \left( \frac{\mu^\dagger}{\mu_{\alpha}^\dagger}\right)^{2-3\vec{\bar{\mu}}/2} \left(\frac{c^{*}\pi\rho_{1}G}{5\gamma}\right)\left[R_\text{c1}(1 + \gamma)\right]^{2},
  \end{equation}
\noindent{where} $R_\text{c1}$ is given by
\begin{equation}
  M_\text{t}^\dagger+m_\text{p}^\dagger = \frac{4\pi\rho_1}{3}R_\text{c1}^3.
\end{equation}

The impact energy is
\begin{equation}
Q^\dagger = \frac{1}{2}\mu^{\dagger}\frac{v_\text{i}^2}{M_\text{t}^\dagger+m_\text{p}^\dagger}.
\end{equation}

\noindent{Following} \citet{Chambers2013}, the mass of the largest remnant for the projectile is calculated as follows

\begin{equation}
M_\text{lr}^\dagger =
\left\{ \begin{array}{lr}
(M_\text{t}^\dagger+m_\text{p}^\dagger) \left(1 - \frac{Q^\dagger}{2Q_\text{RD}^\dagger}\right) & Q^\dagger < 1.8Q_\text{RD}^\dagger \\    
 &\\
0.1 (M_\text{t}^\dagger+m_\text{p}^\dagger) \left(\frac{Q^\dagger}{1.8Q_\text{RD}^\dagger}\right)^{-3/2} & Q^\dagger > 1.8Q_\text{RD}^\dagger.
\end{array}
\right.
\end{equation}

Then, we have to calculate $m_\text{frag} = M_\text{t}^{\dagger} - M_\text{lr}^\dagger$, and distribute it in equal-mass fragments. In the case $m_\text{frag} < M_\text{min}$, we consider the collision as a pure hit-and-run, therefore the masses of both the target and projectile remain unaffected.

\end{enumerate}

This algorithm described in the previous steps is used every time that a collision is detected. However, the computing time required to re-run the collision and classify it is negligible compared to the overall integration time.


\section{Applications - Setup}
\label{section:applications}

In this section, we describe the scenarios of work that will be used in order to test our code.

For the present work, we focus on the study of a dynamical scenario aimed at exploring the sensitivity of the results to the presence of massive perturbers. The methodology to define the physical properties of the planetary embryos to be used in our model is explained as follows.

\subsection{Protoplanetary disk: Model}
\label{sec:model}

Here, we describe the surface density profile that represents a relevant parameter of the properties of the planetary disk. Following \citet{Lynden1974} and \citet{Hartmann1998}, we adopted our model of protoplanetary disk based on the evolution of a thin Keplerian disk only ruled by the gravity of a point-mass central star. The gas-surface density profile $\Sigma_\text{g}(R)$ is given by

\begin{equation}
\Sigma_{\text{g}}(R) = \Sigma_{\text{g}}^{0}\left(\frac{R}{R_{\text{c}}}\right)^{-\gamma} \text{exp}\left[-\left(\frac{R}{R_{\text{c}}}\right)^{2-\gamma}\right],
\label{eq:gas}
\end{equation}
where $R$ is the radial coordinate in the disk mid-plane, $\gamma$ the exponent that represents the surface density gradient, $R_{\text{c}}$ a characteristic radius, and $\Sigma_{\text{g}}^{0}$ a normalization constant.

If we integrate Eq. (\ref{eq:gas}) over the total disk area and assuming axial symmetry, $\Sigma_{\text{g}}^{0}$ can be written as a function of $R_{\text{c}}$, $\gamma$, and the mass of the disk $M_{\text{d}}$ by
\begin{equation}
  \Sigma_{\text{g}}^{0} = (2 - \gamma) \frac{M_\text{d}}{2\pi R_\text{c}^{2}},
\end{equation}
\noindent{where}
\begin{equation}
  M_\text{d} = \int_{0}^{\infty} 2\pi\Sigma_{\text{g}}(R)\text{d}R.
\end{equation}

In the same way, we define a solid-surface density profile $\Sigma_{\text{s}}(R)$ given by
\begin{equation}
\Sigma_{\text{s}}(R) = \Sigma_{\text{s}}^{0}\eta_{\text{ice}}\left(\frac{R}{R_{\text{c}}}\right)^{-\gamma} \text{exp}\left[-\left(\frac{R}{R_{\text{c}}}\right)^{2-\gamma}\right],
\label{eq:solid}
\end{equation} 
where $\Sigma_{\text{s}}^{0}$ is a normalization constant, and $\eta_{\text{ice}}$ a parameter that represents an increase in the amount of solid material due to the condensation of water beyond the snow line. 

In the present study, we assume a central star with a mass $M_{\star}$ = 1 M$_{\odot}$ and a solar metallicity ([Fe/H] = 0). From this, we consider that the relation between the gas and solid surface densities is given by $\Sigma_{\text{s}}^{0} =z_{0}\Sigma_{\text{g}}^{0}$, where $z_{0}$ is the primordial abundance of heavy elements in the sun and has a value of $z_{0} = 0.0153$ \citep{Lodders2009}. Moreover, we adopt a characteristic radius $R_{\text{c}}$ of 25 au and an exponent $\gamma =$ 0.9, which are in agreement with the median values derived from observations of different disks studied by \citet{Andrews2010} in the 1 Myr Ophiuchus star-forming region. Moreover, we adopt a disk mass of $M_\text{d}$ = 0.01 M$_{\odot}$, which is consistent with observational studies developed by several authors, such as \citet{Andrews2010} and \citet{Testi2016}. Finally, we remark that the snow line is assumed to be located at 2.7 au in the present model, according to \citet{Ida2004}. Following \citet{Lodders2003} and \citet{Lodders2009}, we assume that the parameter $\eta_{\text{ice}}$ adopts values of 1 and 2 inside and beyond the snow line, respectively. 

The increase in the amount of solid material due to the condensation of water beyond the snow line produces a radial compositional gradient in the protoplanetary disk. Thus, we propose that the initial fraction of water by mass of the material that compose such a disk is a function of the radial coordinate in the mid-plane $R$ and is given by

\begin{displaymath}
  \text{Water Fraction} =
  \begin{cases}
    10^{-4}\quad \text{$R$ < 2.7 au},\\
    0.50 \quad  \text{ $R$ > 2.7 au}.
  \end{cases}
\end{displaymath}

It is worth noting that this distribution considers that the inner region of the system was populated with water-rich  material from the outer regions during the gaseous phase associated with the evolution of the disk, which is consistent with that proposed by \citet{Raymond2017}.

Finally, we define the habitable zone (HZ) of the system like the region around a star in which a planet could retain liquid water on its surface. In this sense, \citet{Kopparapu2013a} established inner and outer limits for the HZ around stars of different spectral types. In particular, the investigation to be developed in the present work will use the optimistic estimates for a solar-type star derived by those authors, where the inner (outer) edge is located at 0.75 (1.7) au.

In the following section, we describe the necessary parameters for performing the N-body simulations of our research.  

\begin{figure}
 \centering
 \includegraphics[angle=0, width= 0.48\textwidth]{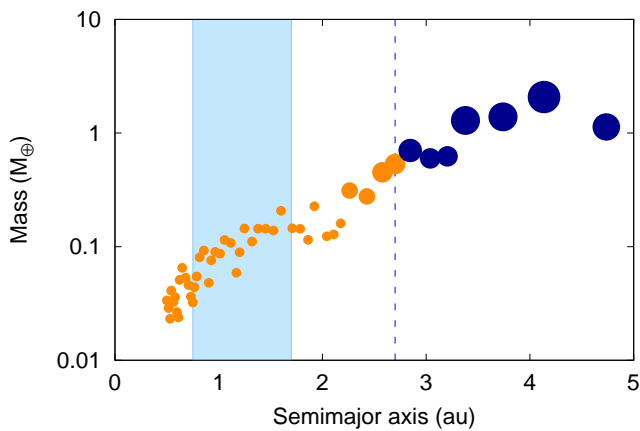}

 \caption{Example of a mass distribution of planetary embryos from one simulation as a function of the initial semimajor axis at the end of the gaseous phase. Note that the size of the points are scaled with the mass of each embryo. Moreover, the different colors illustrate the initial fraction of water by mass for the embryos. In fact, the orange (dark blue) color indicates a fraction of 10$^{-4}$ (0.5) water by mass. The sky-blue shaded region represents the habitable zone for a solar-type star, while the dashed blue line illustrate the snow line assumed in our model.
 }
\label{fig:distr_emb}
\end{figure}
\subsection{N-body simulations: physical and orbital parameters}

For the development of our simulations, we have to specify the physical and orbital parameters of the bodies involved. For that reason, we start specifying the mass distribution for the planetary embryos.

From Eq.\ref{eq:solid}, we can determine the mass distribution once the gas in the disk is fully dissipated. The extension of the region of study is between 0.5 au $< R < $ 5.0 au.

Following \citet{Kokubo2000}, we can determine the mass of an embryo growing in the oligarchic growth regime located at a distance $R$ from the central star like
\begin{equation}
M=2\pi p R\delta R_{\text{H}}\Sigma_{\text{s}}(R),
\label{eq:masaemb}
\end{equation}

\noindent{where} the factor $p$ represents the ratio of the mass of the embryos to the total mass of the system, $\delta R_{\text{H}}$ is the orbital separation of two consecutives planetary embryos of mass $M$ in terms of their mutual Hill radii given by
\begin{equation}
R_{\textrm{H}}=R\left(\frac{2M}{3 M_{\star}}\right)^{\frac{1}{3}},
\label{eq:radiohill}
\end{equation}

{\noindent and} $\delta$ is a parameter randomly chosen following a uniform distribution. The value for $\delta$ is generated each time an embryo's mass and spacing are calculated, ranging between 5 and 10. 

Combining Eqs.\ref{eq:solid} and \ref{eq:radiohill} in Eq.\ref{eq:masaemb}, we can derive a function that relates the mass of each embryo with the distance $R$ as follows
\begin{equation}
  M=\left(2 \pi R^{2} \delta \Sigma_{0\text{s}} \eta_{\text{ice}} p \left(\frac{2}{3 M_{\star}}\right)^{\frac{1}{3}}\left(\frac{R}{R_{\textrm{c}}}\right)^{-\gamma}\text{exp}\left[{-\left(\frac{R}{R_{\text{c}}}\right)}^{2-\gamma}\right]\right)^{\frac{3}{2}}.
  \label{eq:funcion_masa}
\end{equation}

Our scenarios of work only consider planetary embryos for which, we set the value of the factor $p = 1.0$, assuming that there are not planetesimals left at the end of the gaseous phase.

The first embryo is assigned to have an initial semimajor axis $a_1$ = 0.5 au and a mass of $M_1 = 0.036 M_{\oplus}$ (Eq. \ref{eq:funcion_masa}). We calculate the initial semimajor axes and masses for the remaining planetary embryos as follows

\begin{equation}
  a_{i+1} = a_{i} + \delta a_{i} \left(\frac{2M_{i}}{3 M_{\star}}\right)^{\frac{1}{3}},
  \label{eq:distancias}
\end{equation}

\begin{equation}
 M_{i+1} = \left( 2 \pi a_{i+1}^{2}\delta\Sigma_{0\text{s}} \eta_{\text{ice}} p \left(\frac{2}{3M_{\star}}\right)^{\frac{1}{3}}\left(\frac{a_{i+1}}{R_\text{c}}\right)^{-\gamma}\text{exp}\left[-\left(\frac{a_{i+1}}{R_\text{c}}\right)^{2-\gamma}\right]\right)^{\frac{3}{2}}.
\label{eq:masas} 
\end{equation}

We repeat this procedure until the outer limit of our region of study (5.0 au) is reached, obtaining $\sim$ 50 embryos and a total mass distributed of 12.8 M$_{\oplus}$. Figure ~\ref{fig:distr_emb} illustrates the distribution of mass of each embryo as a function of the distance ($R$) from the central star once the gas is fully dissipated. 
As for the initial orbital parameters, we set the values of eccentricity and inclination randomly with a maximum value of 0.02 and 0.5$^\circ$, respectively.
As for the argument of pericenter $\omega$, longitude of ascending node $\Omega$, and the mean anomaly $M$, the values were taken randomly between 0$^{\circ}$ and 360$^{\circ}$.

In order to have a dispersive scenario that favors collisions between planetary embryos, we add two giant planets with masses and physical densities analogous to those of Jupiter and Saturn. Moreover, such planets are assumed to be located on their current orbits, which have semimajor axes of 5.2 au and 9.5 au, and eccentricities of 0.0489 and 0.0565 for Jupiter and Saturn, respectively.

The essential part of this investigation is to study how the treatment of collisions changes the dynamical behavior and the evolution of a planetary system for a given set of initial parameters. We perform a total of 46 simulations using the D3 N-body code, 23 of which take into account a realistic collision prescription and 23 simulations that only consider perfectly inelastic collisions between planetary embryos.

As we mentioned in Sect.~\ref{section:collisional_model}, the value of $M_\text{min}$ is a compromise between computational time and realism. From this, we adopt the value of 0.018 M$_{\oplus}$. A discussion about $M_\text{min}$ and how the variation of this value could affect the overall integration is carried out in Sect.~\ref{section:discusion}. The time-step used in this work is 2 days, which is shorter than 1/40th of the orbital period of the innermost planetary embryo in our simulations. We consider that a embryo collided with the central star if the distance is lower than 0.1 au. This non-realistic stellar radius is so to avoid numerical errors in bodies with very small perihelion. Moreover, a body is considered ejected of the system if the distance is greater than 1000 au. Finally, the integration time for all simulations was of 200 Myr. 

The main results derived from our investigation are presented in the next section. 

\section{Simulation results}
\label{section:results}

In this section, we show a detailed study about the general results obtained from numerical simulations that model the evolution of the dynamical scenario presented in Sect.~\ref{section:applications}. In particular, we carry out a comparative analysis between N-body simulations that incorporate fragmentation and hit-and-run collisions and those that assume that all impacts lead to perfect mergers. This investigation focuses on the physical and dynamical properties of the terrestrial-like planets produced in the HZ of the systems under study, analyzing the role of the fragments in their evolutionary histories.   

\subsection{General analysis}
\label{section:histogram}

The proposed scenario was constructed with the goal of studying the dynamical evolution of a planetary system under formation, subject to the perturbations of two giant planets. As we mentioned in Sect.~\ref{section:applications}, this was carried out by means of a more realistic treatment of the collisions between planetary embryos. This implementation might lead to a difference in the collision history of a planet during its formation. In order to better comprehend this, we analyzed the sensitivity in the collision type distribution.

\begin{figure}[ht!]
 \centering
 \includegraphics[angle=0, width= 0.48\textwidth]{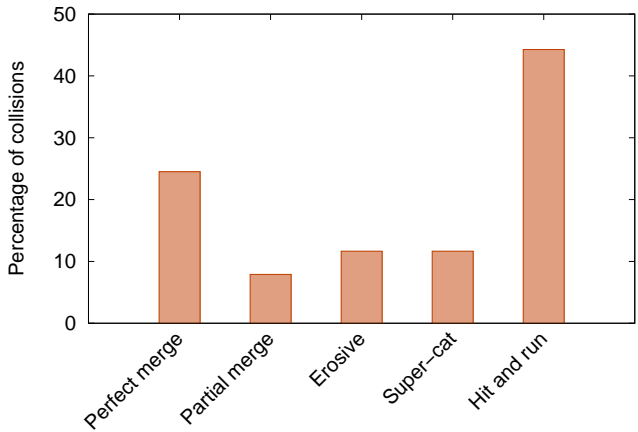}
 \caption{Histogram that shows the percentage of collisions for different regimes implemented in the D3 N-body code for the scenario under study}. 
\label{fig:histograma}
\end{figure}

In Fig. \ref{fig:histograma} we display the percentage of each collision regime for the mentioned scenario. For this figure, only the collisions between massive bodies were taken into account, ignoring those between planetary embryos and fragments, which are always considered as a perfect accretion.

We can observe that the majority of collisions are clumped in two big groups, i.e: perfect merge collisions ($\sim$ 25 $\%$) and hit-and-run encounters ($\sim$ 44 \%). This outcome is consistent with the work of \citet{Kokubo2010}, who found that 49 \% of the collisions between planetary embryos are hit-and-run impacts. Also, our result is consistent with \citet{Chambers2013}, who concluded that, when a refined treatment of collisions is included, about 42 \% of the total number of collisions embryo-embryo results in hit-and-run encounters. In addition to this, we obtained that a considerable percentage of collisions are not either perfect mergers nor hit-and-run impacts. More specifically, this percentage is about $\sim$ 30 \%, including partial accretion, erosive, and super-catastrophic collisions. This wide range of different collisions leads us to consider how these results affect the amount of bodies generated and the overall evolution of the system. 

\begin{figure}[ht!]
 \centering
 \includegraphics[angle=0, width= 0.48\textwidth]{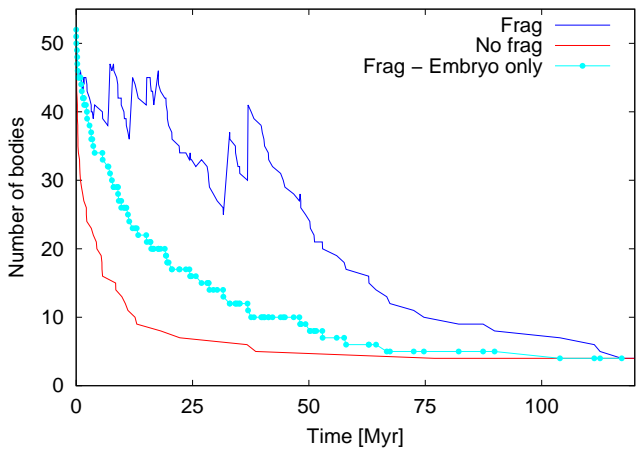}
 \caption{Number of remaining bodies vs integration time. The red curve corresponds to a simulation that lead to perfect mergers. The blue and cyan curves represent a simulation with the realistic collision treatment. In particular, the cyan curve only consider the decrease of planetary embryos. Both curves (red and blue), are representative for each set of simulations.}
\label{fig:nvst_2G}
\end{figure}

\begin{figure*}[ht!]
 \centering
 \includegraphics[angle=0, width=0.49\textwidth]{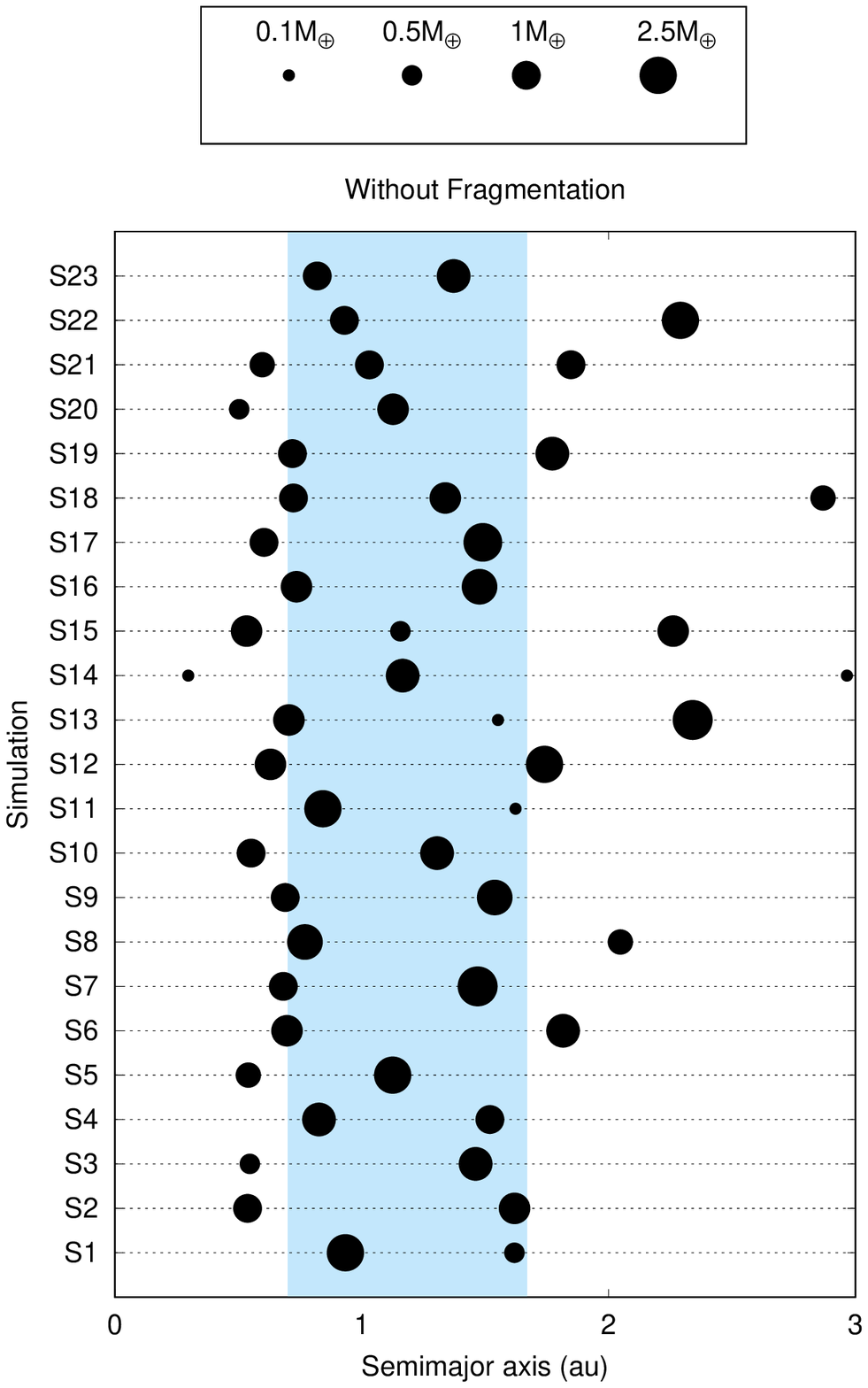} 
 \includegraphics[angle=0, width=0.49\textwidth]{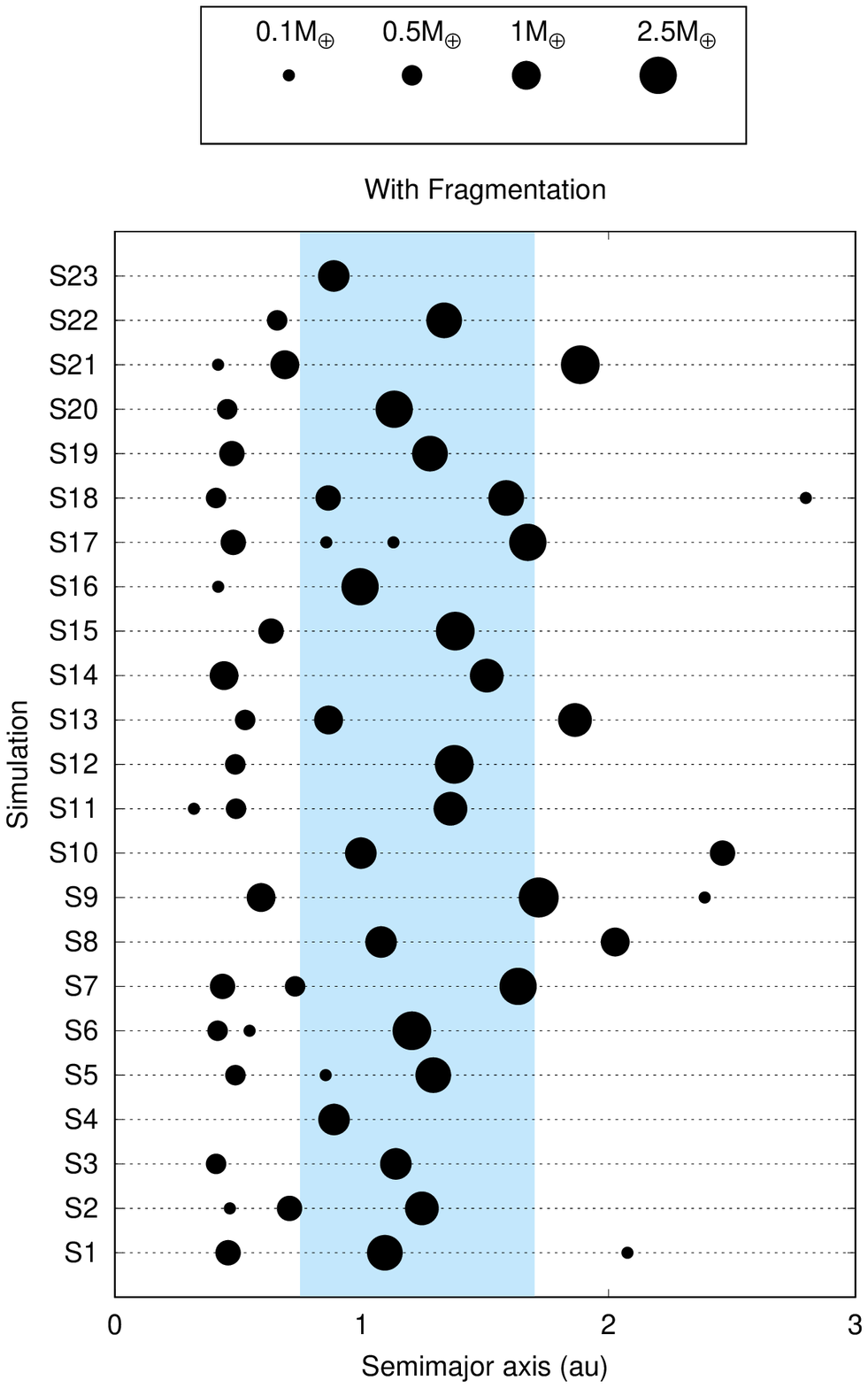} 
 \caption{Final outcomes at 200 Myr of 46 N-body simulations carried out with the D3 code. The left panel illustrates the 23 planetary systems that resulted from runs without fragmentation, while the right panel represents the 23 systems produced from runs that incorporate fragmentation and hit-and-run collisions. The circle size scales with the planet mass according to that indicated in the top region of the figure. The sky-blue area observed in both panels illustrates the habitable zone associated with a solar-type star. \citep{Kopparapu2013a}.
}
\label{fig:tamanios}
\end{figure*}

We can observe in  Fig. \ref{fig:nvst_2G} the total number of bodies as a function of the integration time for a simulation considering only perfectly inelastic collisions (red curve) and another one that assumes a realistic collision treatment (blue curve). The chosen simulations are representatives for each set of runs (fragmentation enabled and fragmentation disabled). The cyan curve shows the decrease of the number of planetary embryos in that simulation represented by the blue curve, but without the fragments generated in each collision.

A natural result is that simulations that include fragmentation take more time for the bodies to decrease in comparison with simulations with perfect accretion. When fragmentation is activated, the time to get half of the initial number of planetary embryos is approximately $\sim$ 11 Myr, while that time decreases down to $\sim$ 5 Myr in runs without fragmentation. This result is a direct consequence of the large number of hit-and-run collisions, which leads to the embryos take more time to collide in comparison with the classical model where this type of encounter is not possible. With this in mind, it is evident that the accretion timescales for systems including a realistic collision prescription are longer than those that only consider perfect mergers, such as is presented in \citet{Chambers2013}.


It is worth noting that the total amount of fragments generated in the simulations range between 50-130. Moreover, we remark that 2-22 fragments were generated from super-catastrophic collisions, while the partial accretions and erosive impacts produced between 1-21 fragments. These numbers could be misleading at first. In fact, a basic assumption is that a super-catastrophic collision should generate more fragments than other impact types, since a minimum of 90 \% of the colliding mass is available to be distributed. However, the number of generated fragments is sensitive not only to the collision regime, but also to the masses of the bodies involved in it. For instance, in the same run, a partial accretion between two bodies with masses of 0.92 M$_{\oplus}$ and 0.88 M$_{\oplus}$ generates 21 fragments, while a super-catastrophic collision between two smaller embryos with masses of 0.16 $M_{\oplus}$ and 0.12 M$_{\oplus}$ only produces 11 fragments. 

In light of these results, we are interested in studying how this realistic collision treatment affects the final architecture of the system and the mass evolution of the resulting planets throughout the entire simulation. Figure \ref{fig:tamanios} illustrates the final architecture for each N-body simulation using the fragmentation improvement model (right panel) and the classical model of perfect accretion (left panel) after 200 Myr of evolution.
The filled black circles represent the final planets and the point size is scaled with their masses, as shown on top of each panel for 0.1 M$_{\oplus}$, 0.5 M$_{\oplus}$, 1 M$_{\oplus}$, and 2.5 M$_{\oplus}$. The sky-blue shaded area indicates the limits of the HZ. We can observe that, in general terms, the number of planets for both sets of runs remains similar, which is consistent with \citet{Chambers2013}, who showed that the final architecture seems insensitive to the collisional model. However, it is important to remark that, although the number of final planets does not differ significantly between the two models, an increase in the number of systems with more than 2 planets is more evident in simulations that incorporate fragmentation and hit-and-run impacts. In fact, 11 runs with fragmentation produce systems with 3 o more planets, while only 5 systems have 3 planets in runs without fragmentation. This result should be interpreted carefully, since a longer integration time associated with simulations with fragmentation might lead to systems with a lower number of final planets.

\begin{figure}[h!]
 \centering
 \includegraphics[angle=0, width= 0.49\textwidth]{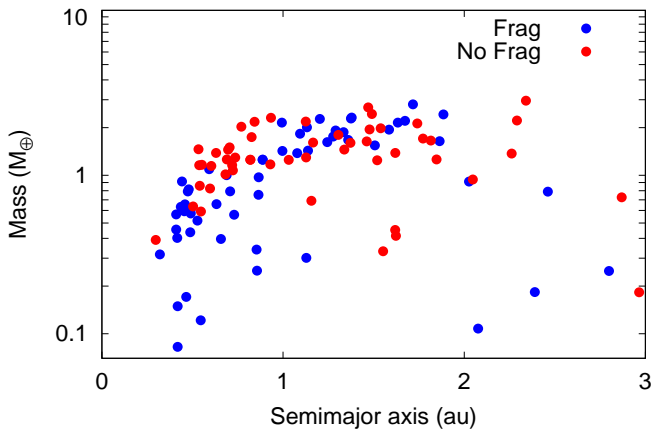}
 \caption{Final mass as a function of the semimajor axis of the planets formed in simulations when fragmentation is activated (blue circles) and without fragmentation (red circles) fragmentation over 200 Myr of evolution.
}
\label{fig:distr_masa_final}
\end{figure}

A distinctive feature observed in almost all simulations of the right panel of Fig.~ \ref{fig:tamanios} indicates the significant existence of low-mass planets around the inner edge of our region of study in comparison with other zones of the system. This feature is not observed in simulations without fragmentation, which are illustrated in the left panel of Fig.~ \ref{fig:tamanios}. This result is also observed in Fig.~ \ref{fig:distr_masa_final}, which represents the final masses of the planets formed in runs with fragmentation (blue points) and without it (red points) as a function of the semimajor axis.

In general terms, Fig.~ \ref{fig:distr_masa_final} shows that the maximum values associated with the overall mass distribution of the final planets formed in simulations without fragmentation tend to be greater than those produced in runs with a realistic collision treatment, reaching 2.95 M$_{\oplus}$. This result should not be surprising since planets formed in simulations with fragmentation could lose mass in each collision while in simulations that assumed that all impacts lead to perfect mergers, each collision that occurs makes the planetary embryo to grow. In the same way, the minimum values observed in the overall mass distribution illustrated in Fig.~  \ref{fig:distr_masa_final} are determined by the blue circles, which are associated with the planets formed in runs with fragmentation and reach 0.08 M$_{\oplus}$. It is important to remark that the difference between the minimum and maximum values of the planetary masses associated with runs with and without fragmentation is enhanced between 0.5 au and 1.0 au. In such a region, we can observe a large concentration of bodies with masses greater than 1 M$_{\oplus}$ for planets formed in simulations without fragmentation, while the planetary masses derived in runs with fragmentation are distributed below those values. 
On the other hand, in the region comprised between 1 au and 2 au, the masses of the largest planets produced in both sets of simulations are comparable. These results are consistent with those observed by \citet{Chambers2013}, who found similar results concerning the final masses of the surviving planets concluding that, when fragmentation is enabled, more low-mass objects are formed, and that the effects of fragmentation over the masses of largest planets are less notorious. 

Finally, Fig.~ \ref{fig:distr_masa_final} shows that there is a significant amount of planets produced in runs with and without fragmentation in the HZ of the system, which is located between 0.75 au and 1.7 au. The study concerning the evolution, survival, and physical properties of such a planets is one of the most important scopes of this work. In the following section, we describe this topic in more detail.

\subsection{Habitable zone planets}

The study of the physical and dynamical properties of the planets formed in the habitable zone (HZ) of the system is very important due to the astrobiological interest of that kind of planets.  

\begin{figure*}[h]
\centering
\includegraphics[angle=0, width= 0.99\textwidth]{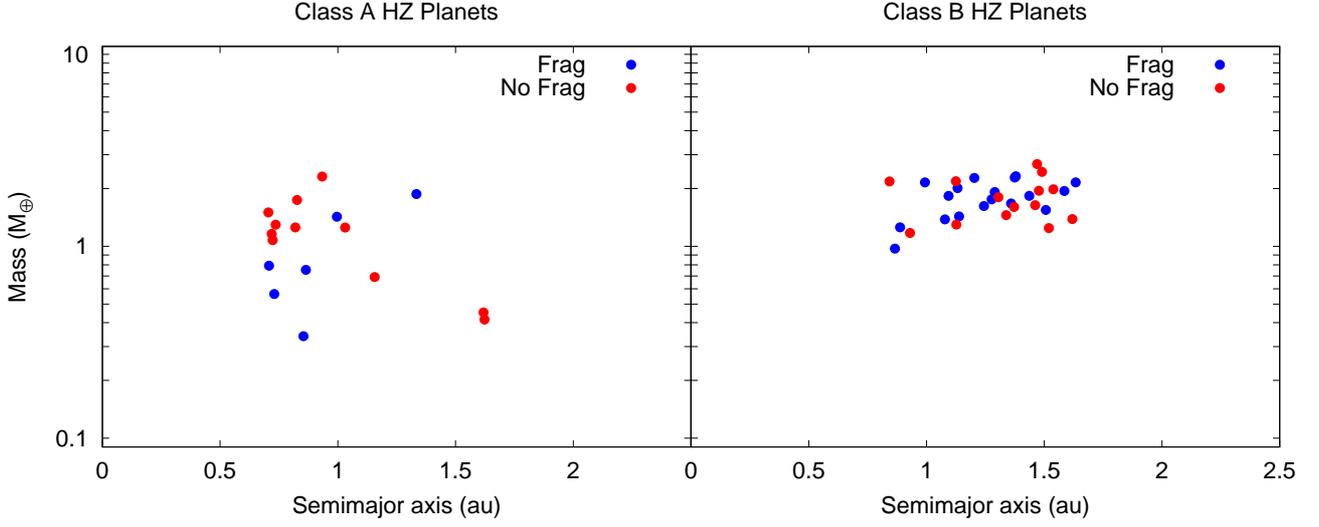}
\caption{Final mass as a function of the semimajor axis of the class A (left panel) and class B (right panel) HZ planets produced in simulations with (blue circles) and without (red circles) fragmentation over 200 Myr of evolution.}
\label{fig:masas_zh}
\end{figure*}

Figure~\ref{fig:masas_zh} shows the mass as a function of the semimajor axis of all planets that survive in the HZ of the systems of study. In particular, the red circles illustrate those planets that result from the N-body simulations that assume that all collisions lead to perfect mergers, while the blue circles show the planets produced from those N-body simulations that incorporate collisional fragmentation and hit-and-run collisions. In general terms, the N-body experiments with fragmentation and hit-and-run collisions form HZ planets with somewhat lower final masses. In fact, the simulations without (with) fragmentation produce HZ planets whose final masses range from 0.42 M$_{\oplus}$ (0.34 M$_{\oplus}$) to 2.68 M$_{\oplus}$ (2.31 M$_{\oplus}$).

It is worth to mention that the planets that are formed and that not belong to the HZ present different physical properties. In particular, the range of masses of the planets in the inner region of the HZ (a $<$ 0.75 au) is between 0.08 M$_\oplus$ - 1.1 M$_\oplus$ in runs with fragmentation while for the simulations where we only consider collisions as perfect mergers the range of mass is ranging between 0.39 M$_\oplus$ - 1.5 M$_\oplus$. As for the outer region of the HZ (1.7 au $<$ a $<$ 3.0 au), the planets present a range of masses of 0.1 M$_\oplus$ - 2.8 M$_\oplus$ in runs with fragmentation and 0.18 M$_\oplus$ - 2.95 M$_\oplus$ for simulations without it. We can observe that the upper limit of the masses increases as we move outside from the central star.

It is important to remark that all our simulations produce two different kinds of planets in the HZ depending on the initial location of their accretion seeds. In simulations without fragmentation the accretion seed is defined as the largest body in each collision \citep{Raymond2009}. For simulations with fragmentation the accretion seed is defined as the largest body in each collision where only one embryo survives. In the case the two embryos survive (pure and erosive hit and run), both of them keep their original seed. According to this, we refer to those planets whose accretion seed starts the simulation inside (beyond) the snow line of the system as class A (class B) HZ planets. From the distribution assumed in Sect.~\ref{sec:model}, the class A HZ planets have very low primordial water contents by mass, while the class B HZ planets have highly significant primordial water contents. Our study produces 6 (11) and 17 (14) class A and class B HZ planets, respectively, in N-body simulations with (without) fragmentation. This information can be found summarized in Tab.~\ref{tab:con_frag}.

The blue and red circles illustrated in the left panel of Fig.~\ref{fig:masas_zh} show the final mass of the class A HZ planets, which result from N-body experiments with and without fragmentation, respectively. Such planets show a broad range of final masses, which reach a minimum and a maximum value of 0.34 M$_{\oplus}$ (0.42 M$_{\oplus}$) and 1.87 M$_{\oplus}$ (2.30 M$_{\oplus}$) in N-body simulations with (without) fragmentation, respectively. According to this, the minimum and maximum values of the final mass of the class A HZ planets resulting from N-body runs with fragmentation are about 20 \% smaller than those produced without fragmentation. 

In order to understand such differences in the final masses, we carry out an analysis of the collisions involved in the evolutionary history of the class A HZ planets formed in N-body experiments without fragmentation and in those that incorporate collisional fragmentation and hit-and-run collisions. 

In runs without fragmentation, the 11 class A HZ planets undergo a total of 70 perfect mergers. While one of them does not experience any collision during 200 Myr, and other 2 of such planets undergo between 1 and 2 perfect mergers, the other 8 class A HZ planets experience more than 7 perfectly inelastic collisions, which lead to significant values in the relative growth of their masses respect to the initial values. 

It is important to mention that the collisions that are taken into account are only those with the surviving embryos. We do not count the possible collisions that the projectiles may had.

\begin{figure*}[h]
\centering
\includegraphics[angle=0, width= 0.45\textwidth]{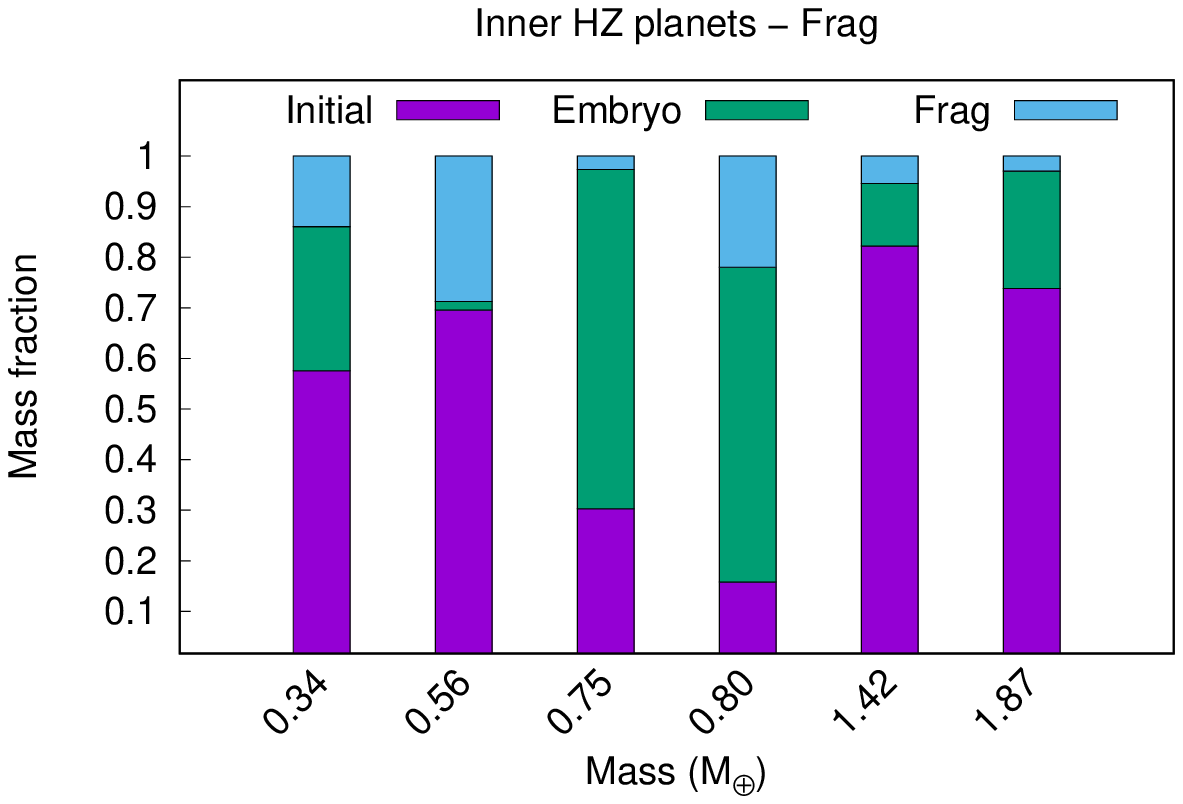}
\includegraphics[angle=0, width= 0.45\textwidth]{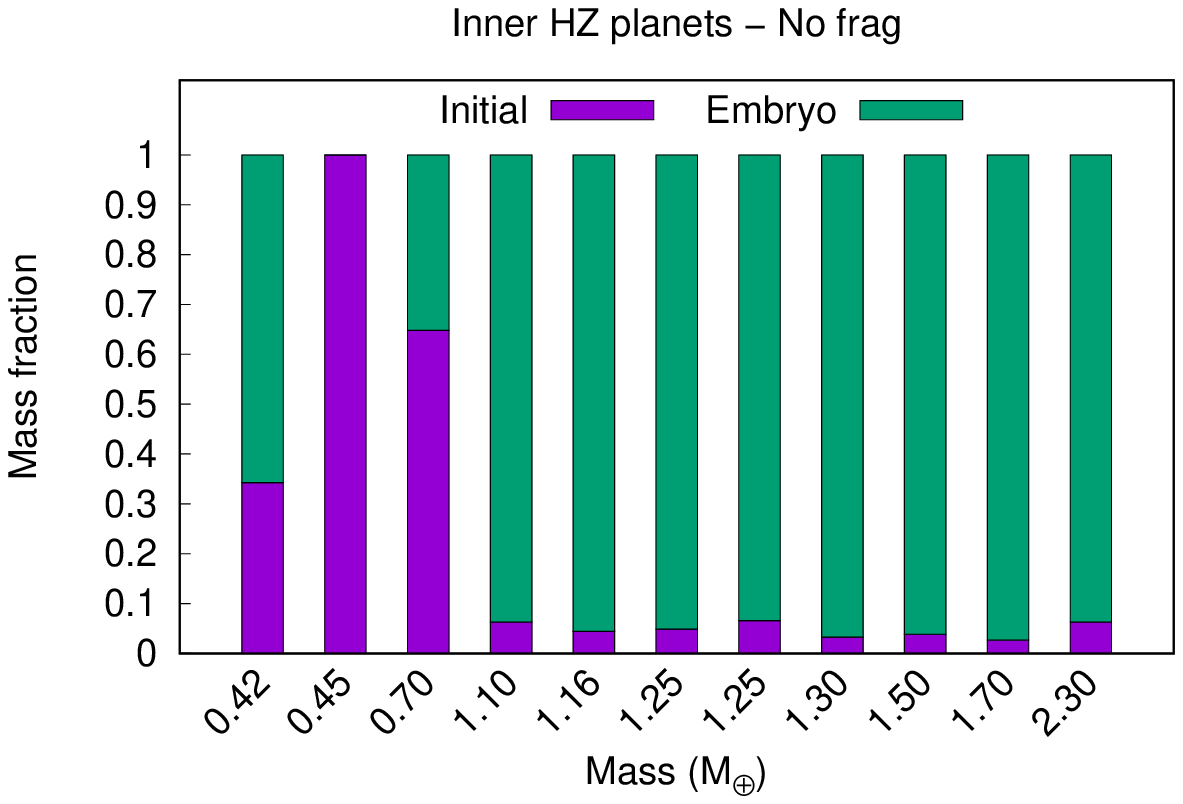}\\
\includegraphics[angle=0, width= 0.45\textwidth]{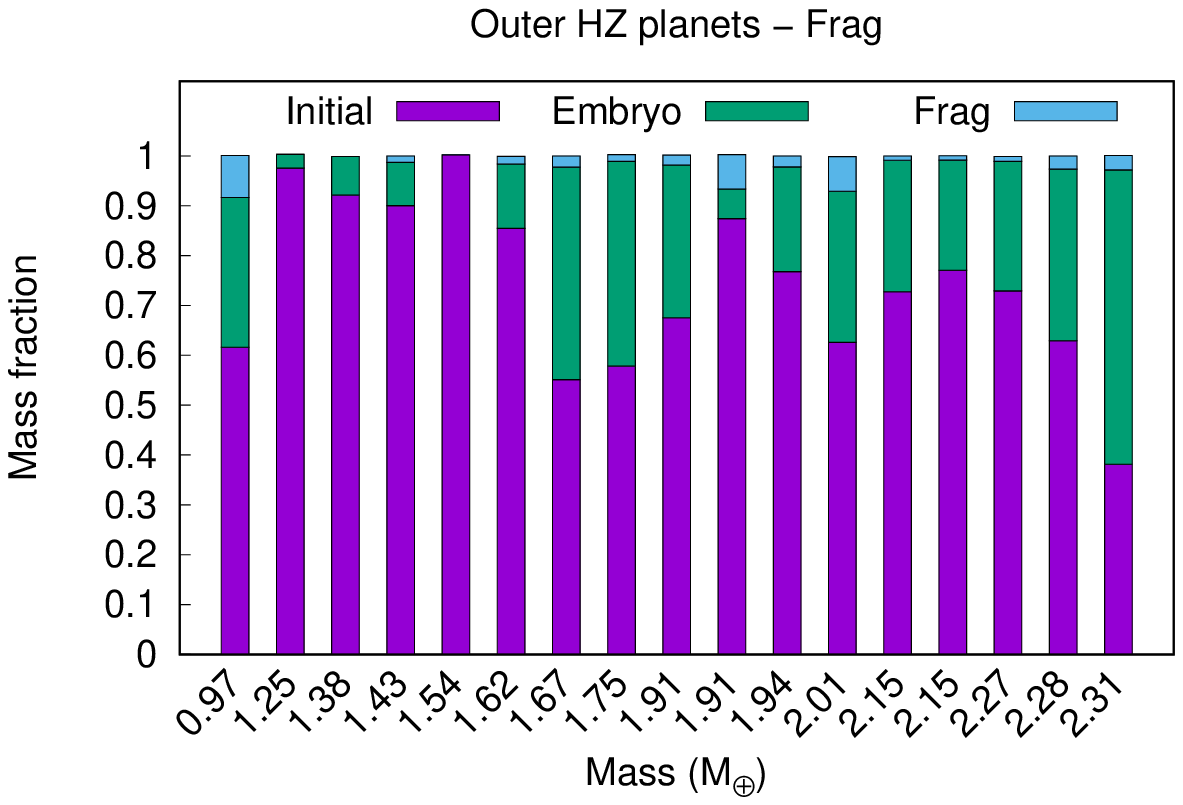}
\includegraphics[angle=0, width= 0.45\textwidth]{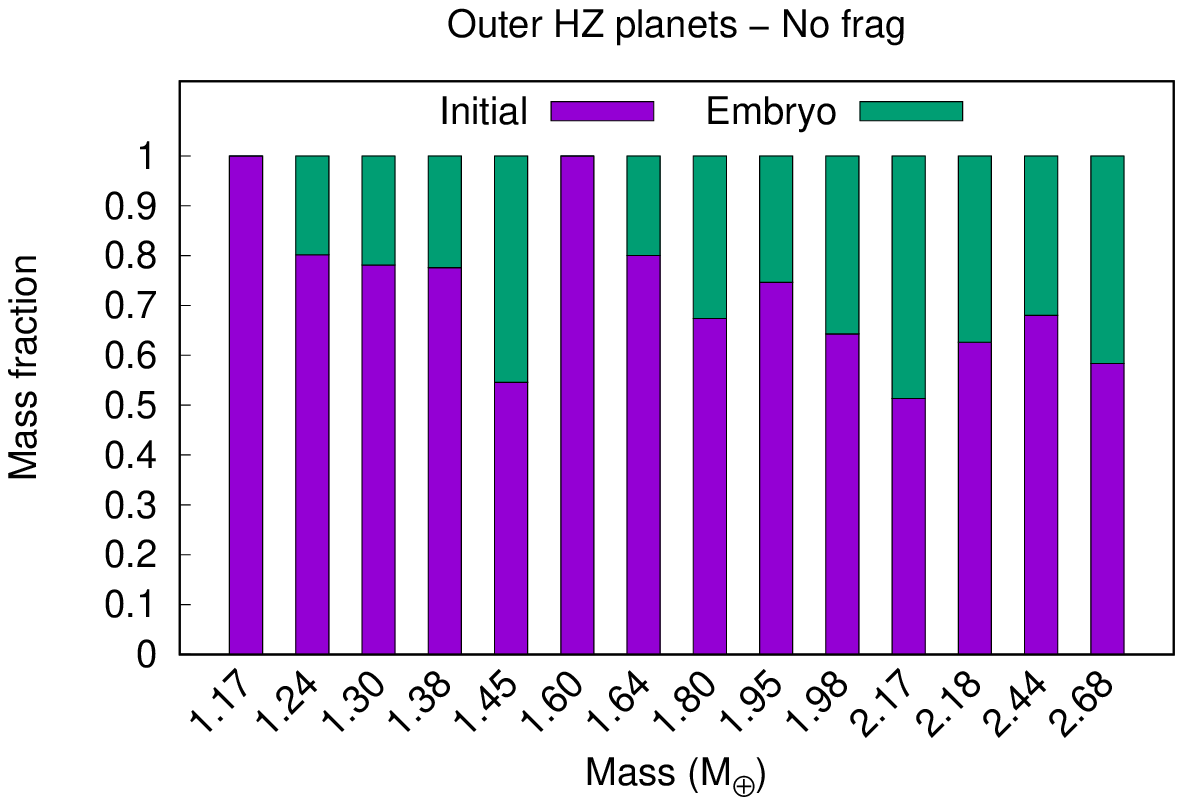}
\caption{Distribution of the final mass for the surviving planets in the HZ. The violet bars represent the mass fraction corresponding to the initial mass of the body. The green and sky-blue bars represent the mass fraction contribution due to partial (or total) accretion of embryos and fragments, respectively. The top panels display the results for the class A HZ planets for Frag (left) and No Frag (right) simulations. The bottom panels are analogous to the top panels but for the class B HZ planets. The height of each bar is normalized to the final mass of each planet.}
\label{fig:fracciones_masa_final}
\end{figure*}

This result can be observed in the top and right panel of Fig.~\ref{fig:fracciones_masa_final}, which illustrates the fraction contributed by the initial mass (violet) and perfect mergers with embryos (green) to the final mass of the class A HZ planets. According to this, the 8 most massive class A HZ planets with $M \gtrsim$ 1 M$_{\oplus}$ produced in N-body simulations without fragmentation reach more than 90 \% of their final masses from perfectly inelastic collisions with planetary embryos over 200 Myr of evolution.

In runs with fragmentation, the 6 class A HZ planets produced in such simulations experience a total of 57 collisions, of which 9 are perfect mergers with planetary embryos, 27 perfect mergers with fragments, 6 partial accretions, 2 erosive impacts, 9 hit-and-run, and 4 erosive hit-and-run. According to this, the number of partial accretions and perfect mergers with planetary embryos is relatively low in the evolutionary history of the class A HZ planets in comparison with the number of perfect mergers with generated fragments. In fact, partial accretions and perfect mergers with planetary embryos (fragments) represent about 10 \% and 16 \% (47 \%) of the total number of collisions experienced by those planets, respectively. However, it is important to remark that the perfect mergers with generated fragments play a secondary role in the growth of the class A HZ planets. In fact, such as the top and left panel of Fig.~\ref{fig:fracciones_masa_final} shows through sky-blue boxes, the fragments contribute with less than 30 \% to the final mass of those planets. An important difference observed respect to the simulations without fragmentation is associated with the relative growth of the class A HZ planets more massive than 1 M$_{\oplus}$. In fact, such as the top and left panel of Fig.~\ref{fig:fracciones_masa_final} shows through violet boxes, the initial masses of such planets represent more than 70 \% of their final masses, which indicates that their relative growths are significantly less than those associated with the class A HZ planets with $M \gtrsim$ 1 M$_{\oplus}$ produced in runs without fragmentation. Our results seem to suggest that the most massive class A HZ planets in simulations with fragmentation require relatively high initial masses.

\begin{figure*}[h]
\centering
\includegraphics[angle=0, width=0.45\textwidth]{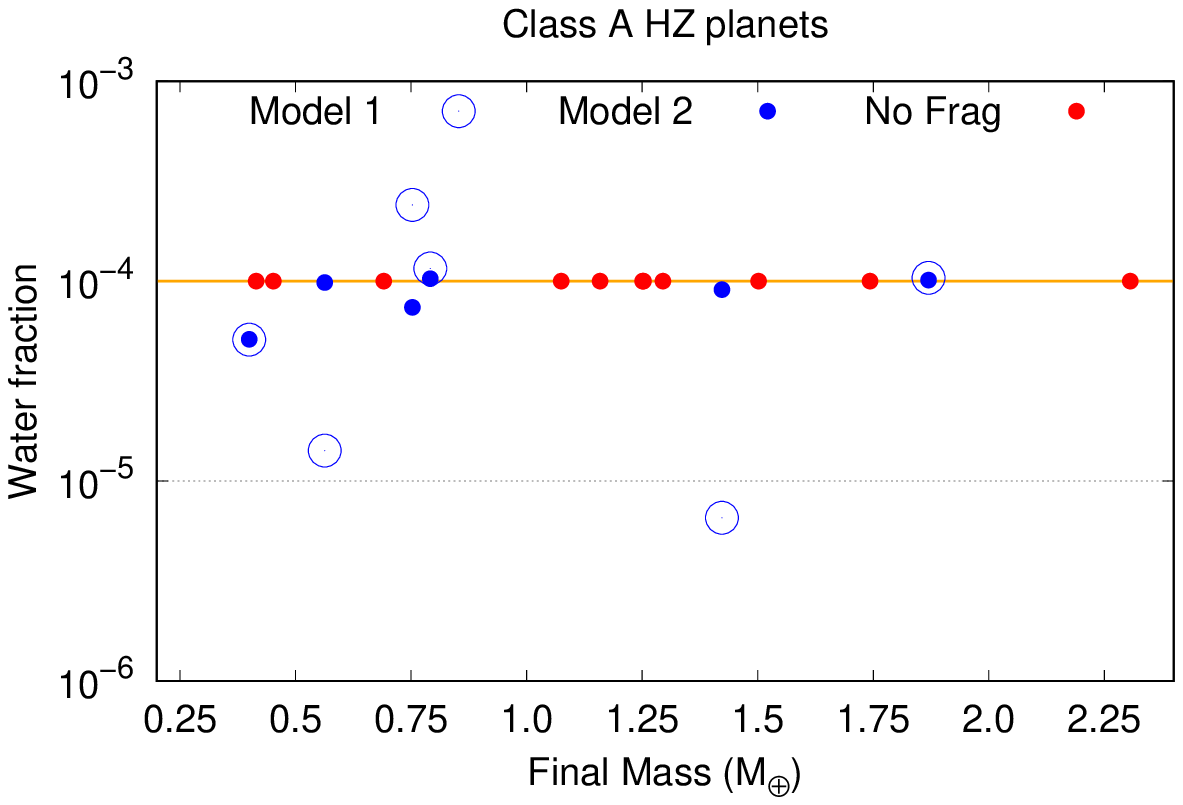}
\includegraphics[angle=0, width= 0.45\textwidth]{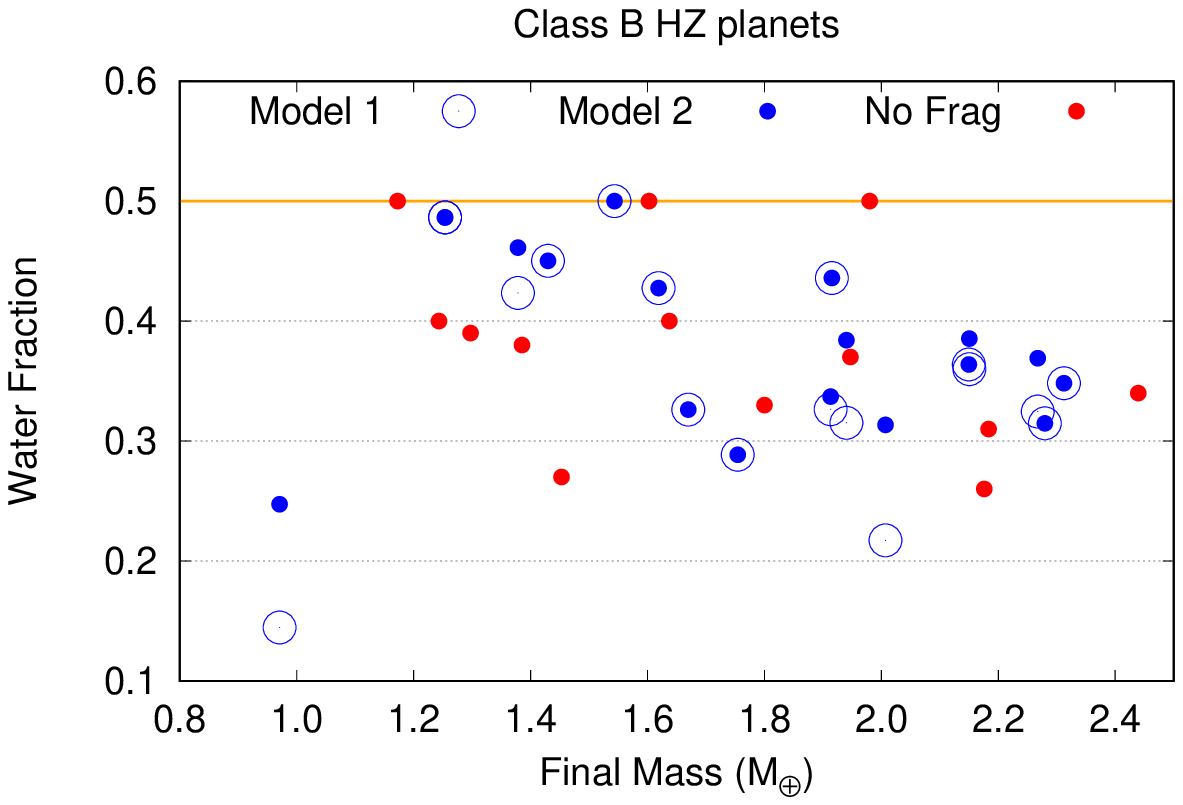}
\caption{Final fraction of water by mass as a function of the final mass of the class A HZ planets (left panel) and class B HZ planets (right panel). The red filled circles illustrate the values associated with those planets formed in simulations without fragmentation, while the open and filled blue circles represent the values obtained from Models 1 and 2 in runs with fragmentation, respectively. In both panels, the black solid line represent the initial water fraction for class A and class B HZ planets.}
\label{fig:fracciones_agua_internos_todos_juntos}
\end{figure*}

\begin{figure*}[h]
\centering
\includegraphics[angle=0, width= 0.45\textwidth]{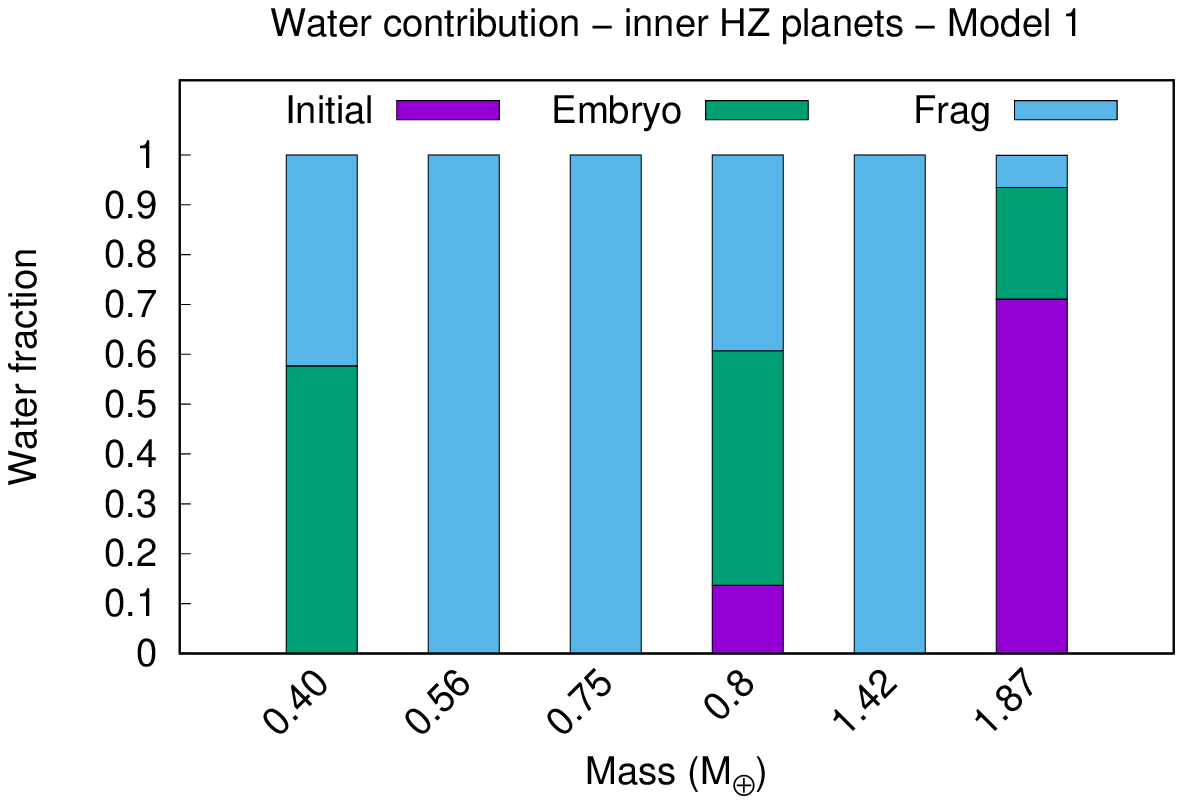}
\includegraphics[angle=0, width= 0.45\textwidth]{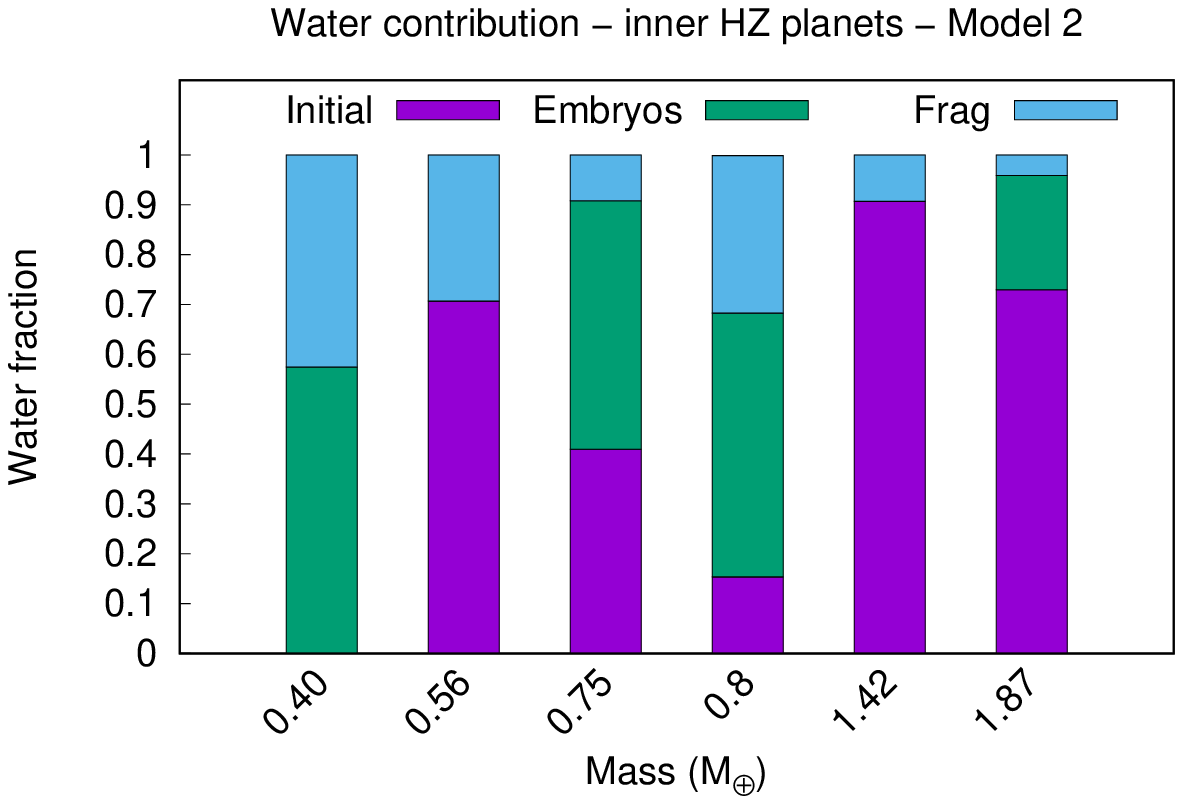}
\caption{Same as Fig. \ref{fig:fracciones_masa_final} for the water contribution. The color code is the same as in the mentioned figure. The panels represent the water contribution to the class A HZ planets using Model 1 (left panel) and 2 (right panel), respectively.}
\label{fig:fracciones_agua_internos_dos_paneles}
\end{figure*}

At the same way, the blue and red circles represented in the right panel of Fig.~\ref{fig:masas_zh} illustrate the final mass of the class B HZ planets, which are formed from N-body experiments with and without fragmentation, respectively. In this case, the resulting planets show a more narrow range of final masses. In fact, the less and most massive class B HZ planet produced in runs with (without) fragmentation has 0.97 M$_{\oplus}$ (1.17 M$_{\oplus}$) and 2.31 M$_{\oplus}$ (2.68 M$_{\oplus}$), respectively. In this case, the minimum and maximum values of the final mass of the class B HZ planets produced from N-body runs with fragmentation are about 15 \% smaller than those obtained without fragmentation. 
Again, it is necessary to carry out an analysis of the collisions undergone by the class B HZ planets produced in our N-body simulations. In runs without fragmentation, the 14 class B HZ planets experience a total of 27 perfect mergers with embryos. While 2 of them do not undergo any collision over 200 Myr of evolution, the other 12 class B HZ planets experience between 1 and 4 perfectly inelastic collisions. According to that observed in the bottom and right panel of Fig.~\ref{fig:fracciones_masa_final} through violet boxes, the initial masses of the class B HZ planets produced in runs without fragmentation represent more than 50 \% of their final masses. From this, the relative growth of such planets is less significant than that experienced by the class A HZ planets with $M \gtrsim$ 1 M$_{\oplus}$ formed in simulations without fragmentation. 

\begin{table*}[h!]
  \begin{center}
   
    \begin{tabular}{|c|c|c|c|c|c|c|c|c|} 
    \hline
      \textbf{Frag} & \textbf{\# HZ Planets} & \textbf{Class A} & \textbf{Class B} & \textbf{Masses (\textbf{Class A})} & \textbf{Masses (\textbf{Class B})} & \textbf{Wt  fr. Class A} & \textbf{Wt fr. Class B)}\\
      \hline
     On  & 23 & 6  & 17 & 0.34 - 1.87 [M$_{\oplus}$]&  0.97 - 2.31 [M$_{\oplus}$] & 
     6.5  $\times$ 10$^{-6}$ - 1.0  $\times$ 10$^{-4}$ & 0.15 - 0.5 \\
     Off & 26 & 11 & 14 & 0.42 - 2.30 [M$_{\oplus}$] &  1.17 - 2.68 [M$_{\oplus}$] & 1.0 $\times$ 10$^{-4}$ & 0.26 - 0.5 \\
      \hline
    \end{tabular}
    \vspace{5mm}
       \caption{ Summarize of amount of bodies formed for class A and class B HZ planets and their range of masses and final water fraction, in runs with and without fragmentation.}\label{tab:con_frag}
    
  \end{center}
\end{table*}

\begin{table*}[h!]
\begin{center}
   
    \begin{tabular}{|c|c|c|c|c|c|c|} 
    \hline
      \textbf{HZ planets} & \textbf{Perfect Merge} & \textbf{Partial} & \textbf{Erosive} & \textbf{Super-Cat} & \textbf{hit and run}\\
      \hline
     \textbf{class A} & 36  & 6 & 2  &  0 & 13            \\
     \textbf{class B} & 67 & 9  & 1  &  0 & 16           \\
      \hline
    \end{tabular}
      \vspace{5mm}
   \caption{ Total amount and type of collisions ocurred for class A and class B HZ planets.}
   \label{tab:con_frag2}
  \end{center}
\end{table*}

In runs with fragmentation, the 17 class B HZ planets formed in such simulations undergo a total of 93 collisions, of which 33 are perfect mergers with planetary embryos, 34 perfect mergers with fragments, 9 partial accretions, 1 erosive impact, 6 hit-and-run, and 10 erosive hit-and-run. In Tab.~\ref{tab:con_frag2} can observe this information summarized. In this case, the partial accretions and the perfect mergers with planetary embryos (fragments) represent about 10 \% and 35 \% (37 \%) of the total number of collisions undergone by the class B HZ planets, respectively. As commented on the class A HZ planets, the contribution of the generated fragments to the final mass of the class B HZ planets is also secondary in this case, since they provide less than 10 \%, which is shown in the bottom and left panel of Fig.~\ref{fig:fracciones_masa_final} through sky-blue boxes. It is worth mentioning that, while 3 of the class B HZ planets formed in runs with fragmentation do not undergo any perfect merger with an embryo over 200 Myr of evolution, the other 14 experience between 1 and 5 of such perfectly inelastic collisions, which is comparable to that described for the class B HZ planets produced in simulations without fragmentation. From this and the low number of collisions that lead to mass losses, the fraction contributed by the initial mass and perfect mergers with embryos to the final mass of the class B HZ planets is similar in simulations with and without fragmentation. This comparative analysis can be observed from the results illustrated in the bottom panels of Fig.~\ref{fig:fracciones_masa_final}, where the contribution provided by the initial mass and the perfect mergers with embryos is represented through violet and green boxes, respectively. It is worth to mention that for the class A HZ planets in simulations with and without fragmentation their feeding zones are fully contained in the region inner to the snowline. While for class B HZ planets, their feeding zones are mainly contained in the outer region of the snowline, although there's a minor contribution from the inner region. This result, will affect to the final water content of the final bodies.

The final water content of the class A and class B HZ planets plays a key role in order to understand their potential astrobiological interest. To quantify the final fraction of water by mass of such planets, we track their collisional histories throughout the entire evolution. For N-body simulations without fragmentation, the treatment is very simple since all collisions are assumed to result in perfect mergers, which conserve the mass and the water content of the interacting bodies. For N-body experiments that incorporate collisional fragmentation and hit-and-run collisions, the treatment is more complex since we must specify the amount of water acquired by the largest remnant and the fragments generated in each event. To do this, we define the following two models, which are based on prescriptions proposed by \citet{Marcus2010}:
\begin{itemize}
\item Model 1: the fraction of water of the largest remnant of each collision is determined by assuming that the mass that escapes is water from the projectile (first) and the target (second), and then rocky material from such bodies in the same order, 
\item Model 2: the fraction of water of the largest remnant of each collision is determined by assuming that the mass that escapes is water (first) and rocky material (second) from the projectile, and then water and rocky material from the target in the same order. 
\end{itemize}
It is important to remark that Model 1 leads to a largest remnant with a less fraction of water by mass than that derived from Model 2. However, it is worth mentioning that Model 1 produces a distribution of fragments with a greater amount of water than that generated from Model 2. Thus, the final water content of the HZ planets will strongly depend on the class of impacts that they undergo throughout the entire evolution.

The left panel of Fig.~\ref{fig:fracciones_agua_internos_todos_juntos} illustrates the final fraction of water by mass as a function of the final mass of the class A HZ planets formed in N-body simulations with and without fragmentation, which are represented by blue and red circles, respectively. In particular, the open and filled blue circles show the results derived using Models 1 and 2 to determine the water transfer in runs with fragmentation, respectively. 

As the reader can see in the left panel of such a figure, all class A HZ planets produced in N-body runs without fragmentation conserve the initial value of 10$^{-4}$ associated with the fraction of water by mass over 200 Myr of evolution, since all collisions are assumed to result in perfect mergers and their feeding zones are only associated with the region of the system inside the snow line. Moreover, it is important to remark that the primordial water contents of 8 of the 11 class A HZ planets formed in runs without fragmentation represent less than 10 \% of their final water contents. This can be inferred from the top and right panel of Fig.~\ref{fig:fracciones_masa_final}. In fact, since the feeding zones of such planets are restricted to the region of the system inside the snow line, the contributions to the mass fraction normalized to the final mass are equivalent to the contributions to the water fraction normalized to the final water content. Thus, in general terms, in our N-body experiments without fragmentation, the main source of water of the class A HZ planets are through collisions with embryos, instead of from its primordial water content. 

The analysis is more complex for the class A HZ planets that result from simulations with fragmentation. In such a case, the final fractions of water of the class A HZ planets depend on the model adopted to track the evolution of water. In fact, it is important to remark that the perfect mergers with generated fragments represent about 47 \% of the total number of collisions experienced by the class A HZ planets. Thus, while the fragments play a secondary role in the final masses of such planets, they can be important in order to determine the final fraction of water of such planets.    

As we have already mentioned above, the left panel of Fig.~\ref{fig:fracciones_agua_internos_todos_juntos} particularly illustrates the final fraction of water by mass as a function of the final mass of the class A HZ planets formed in simulations with fragmentation. According to this, the final fractions of water by mass derived from Model 1 (open blue circles) show significant differences respect to those obtained using Model 2 (filled blue circles). In fact, Model 1 offers a wide range of results. Indeed, such a model can lead to slight increases of $\sim$ 2.4 $\times$ 10$^{-4}$ and significant decreases of $\sim$ 6.5 $\times$ 10$^{-6}$ in the final fraction of water by mass of the class A HZ planets respect to their initial fraction of water of 10$^{-4}$. Moreover, Model 1 can also keep the final fraction of water by mass of some of such planets close to the initial value of 10$^{-4}$. On the contrary, Model 2 shows more conservative results concerning the final fraction of water. From such a model, 4 of 6 class A HZ planets show a final fraction of water by mass very close to the initial value of 10$^{-4}$. The other 2 planets undergo a slight decreases of the fraction of water by mass over 200 Myr respect to the initial fraction, reaching a minimum final value of $\sim$ 5 $\times$ 10$^{-5}$, which is associated with the less massive class A HZ planet. 

According to this, in general terms, the water distribution between the largest remnant and the generated fragments in a given collision proposed by Model 2 offers results comparable to those obtained in simulations without fragmentation concerning the final fraction of water by mass of the class A HZ planets. In this sense, Model 1 shows more diverse results about the final fraction of water by mass of such planets, leading to slight increases, significant decreases, and even similar fractions respect to those derived in runs without fragmentation. 

It is very interesting to determine the main source of water of the class A HZ planets formed in simulations with fragmentation for each of the two models proposed to track the evolution of water. In order to analyze this, the left and right panels of Fig.~\ref{fig:fracciones_agua_internos_dos_paneles} illustrate the contribution of the primordial water content (violet), and that provided by collisions with planetary embryos (green) and generated fragments (sky-blue) to the final fraction of water by mass of the class A HZ planets, which is derived making use of Models 1 and 2, respectively. 

From this, it is possible to observe that, in general terms, the role of the generated fragments is very important in the final fraction of water by mass of the class A HZ planets when Model 1 is used. In fact, the fragments represent the main source of water in 3 of 6 of such planets, providing the totality of their final contents. In 2 of 6 class A HZ planets, about 40 \% of their final water contents is acquired by perfect mergers with fragments. These examples are very important since they extends the significance of the fragments in cases where the planets receive a relevant contribution of water from collisions with planetary embryos. Finally, it is worth noting that the fragments just play a secondary role in the final water content of the most massive class A HZ planet, in which the primordial water content represents about 70 \% of its final content. Beyond this, in general terms, it is necessary to remark that the primordial water content does not play a relevant role in the final water content of the class A HZ planets formed in runs with fragmentation when Model 1 is adopted. 

The conclusion concerning the contributions of the primordial water, planetary embryos, and generated fragments to the final water content of the class A HZ planets produced in simulations with fragmentation is very different when Model 2 is used. In this particular case, the general result shows that the generated fragments play a minor role in the final fraction of water by mass of such planets in comparison with that described using Model 1. In fact, this is clearly observed if we focuses the analysis on those 3 class A HZ planets that acquire the totality of their final water contents by perfect mergers with fragments using Model 1. If Model 2 is adopted, the fragments just provide about 10 \% (30 \%) of the final water content of 2 (1) of those 3 class A HZ planets. In general terms, it is important to note that the primordial water content has a relevant contribution to the final water content in 4 of 6 class A HZ planets, which represents another important difference with that derived from Model 1.   

Our results show that the main source of water and the final fraction of water by mass of the class A HZ planets produced in simulations with fragmentation are very sensitive to the model adopted to determine the distribution of water between the largest remnant and the generated fragments in a collision. According to this, it is very necessary to specify a realistic model of volatile transport in order to determine in detail the water abundances of the formed terrestrial-like planets from N-body experiments.  

The situation is very different when the final water content of the class B HZ planets is analyzed. In fact, such as the right panel of Fig.~\ref{fig:fracciones_agua_internos_todos_juntos} illustrates, the final fractions of water by mass of the class B HZ planets produced in runs with fragmentation are represented by open (filled) blue circles and range from 15\% (25\%) to 50\% (50\%) when Model 1 (2) is used. From this, we derive two important conclusions. On the one hand, in general terms, the final fraction of water by mass of the class B HZ planets does not strongly depend on the model adopted to determine the distribution of water in a given collision. Indeed, 12 of 17 of such planets reach almost the same final fraction of water by mass with both models. On the other hand, our results show that the collisional fragmentation is not a barrier to the formation and survival of water worlds in the HZ of the system. Another interesting result represented in the right panel of Fig.~\ref{fig:fracciones_agua_internos_todos_juntos} indicates that the final fractions of water by mass of the class B HZ planets produced in runs without fragmentation (red circles) show similar values to those obtained in simulations with fragmentation. This point is very important since the results of simpler N-body simulations without fragmentation concerning the water content of the so-called water worlds should provide a good first approximation to the real data.       

\section{Discussion and Conclusions}
\label{section:discusion}

In the present research, we analyze the formation and evolution of terrestrial-like planets around solar-type stars, in systems that host two giant planets with physical properties and orbital parameters similar to those associated with Jupiter and Saturn. In particular, our investigation is based on a comparative analysis of 46 N-body simulations, which are carried out using an own numerical code called D3 N-body code. This numerical tool incorporates fragmentation and hit-and-run collisions according to the prescriptions described in the works developed by \citet{Leinhardt2012}, \citet{Genda2012}, \citet{Chambers2013}, and \citet{Mustill2018}. To develop our study, 23 out of 46 N-body simulations are carried out assuming a realistic treatment for collisions, while the other 23 N-body experiments are developed considering that all collisions lead to perfect mergers. In particular, our study focuses on the physical and dynamical properties of the terrestrial-like planets and water delivery in the Habitable Zone (HZ) of the system. 

It is worth noting that, unlike previous work like \citet{Chambers2013}, \citet{Quintana2016}, and \citet{Wallace2017} in our N-body simulations we did not include planetesimals for studying the evolution of a planetary system. Our protoplanetary disks are composed only by planetary embryos. We understand that, in order to have real initial conditions and determining the mass ratio between planetesimals and embryos, it is necessary to have a detailed evolution of the gaseous stage. Here a simple model is proposed. In this model we can compare the physical properties of terrestrial-like planets in N-body simulations with and without fragmentation.

It is important to remark that our comparative analysis between simulations with and without fragmentation derives results consistent with those obtained by \citet{Chambers2013}, concerning the general percentage of the different collisional regimes, temporal evolution of the number of objects of the system, final mass of the terrestrial-like planets, and final planetary architecture.

The physical properties of the terrestrial-like planets formed in the HZ of systems that result from simulations with and without fragmentation show significant differences in several aspects. In particular, our conclusions vary for class A and class B HZ planets, which show accretion seeds initially located inside and beyond the snow line, respectively.  

In general terms, the planetary embryos represent the main source of mass and water of the class A HZ planets produced in runs without fragmentation. In particular, the class A HZ planets more massive than 1 M$_{\oplus}$ experience very significant increases of mass and water respect to their initial values. Beyond this, the final fraction of water of all class A HZ planets formed in simulations without fragmentation is equal to their initial value of 10$^{-4}$, since all collisions are treated as perfect mergers and their feeding zones do not include water-rich embryos, initially located beyond the snow line.   

As for the simulations that incorporate fragmentation and hit-and-run collisions, the generated fragments play a secondary role in the mass of the class A HZ planets, providing less than about 30 \% of their final value. In particular, the class A HZ planets more massive than 1 M$_{\oplus}$ undergo small increases of mass respect to their initial values, since the primordial mass contributes with more than 70 \% to the final mass. 

It is important to remark that the role of the fragments in the final water content of the class A HZ planets, as well as their final fractions of water, strongly depend on the model adopted to determine the distribution of water in a given particular collision. On the one hand, if water from the projectile and the target escape first, the generated fragments play a very important role in the final water content of the class A HZ planets, while the primordial water content does not represent a significant fraction of the final content. Moreover, the final fraction of water of such planets covers a wide range of values, which show slight increases, significant decreases, and even very low deviations respect to the initial fraction of 10$^{-4}$. On the other hand, if water and rocky material from the projectile escape first, the role of the fragments in the final water content of the class A HZ planets is less significant, while the contribution of the primordial content of water plays a primary role. Moreover, the final fractions of water of such planets show values close to the initial fraction of 10$^{-4}$, which is comparable with that obtained for the class A HZ planets formed in runs without fragmentation. 

As for the class B HZ planets, the conclusions are very different. First, the physical properties of such planets formed in runs with and without fragmentation are very similar. In fact, it is important to remark that the contribution of the generated fragments to the final final water content of the class B HZ planets is negligible regardless of the model adopted to distribute the water after a given collision. This can be attributted to the fact that there are no erosive impacts in water rich embryos, leaving water-rich fragments that can contribute significantly to the final water content of class B HZ planets. On the other hand, in regards to the final mass of the planets the generated fragments play a secondary role contributing in some cases up to 10\% of the final mass of the planet. The most important result derived for this kind of planets indicates that the incorporation of the fragmentation in the collisional algorithm does not prevent the formation and survival of water worlds in the HZ of our systems.  

It is worth noting that the model that we have adopted to carry out our investigation has some limitations that should be considered. First, we remark that the mass of each fragment has the same constant value in all our simulations. The selection of a smaller value for such a mass would lead to more generated fragments per collision, which may affect our results. In this sense, \citet{Wallace2017} analyzed the role of the fragmentation in the rocky planet formation at small distances from the central star, and investigated the dependence of their results on the choice of the mass of fragments in the collisional algorithm. While these authors did not find significant discrepancies in N-body simulations developed using different values for the fragment mass, we consider that it would be interesting to investigate the role of such a mass value in the final physical properties of the planets formed in the HZ of the system. 

As for the models adopted from Marcus et al. 2010 for determine the water content of the final planets have to be looked carefully. We use two models for mantle stripping in a differentiated body for modelling the water loss in a collision between two planetary embryos. Although this is true for water-rich bodies, we have to be careful with bodies with a low percentage of water. One caveat of this model is that, in reality, the water may not actually appear as an icy-mantle, but rather inside the silicate mantle and atmosphere. In consequence, the model would no longer be valid and the results may change abruptly. Mantle stripping models that could be used for low or high water content have to be specified in order to achieve a good description of the final water contents. Creating more complex models for water loss by mantle stripping is in development for including it in future works.

On the other hand, we consider that the treatment of the smaller collisional fragments is a point of study to be checked in future works. In fact, such as is described in \citet{Chambers2013} and \citet{Wallace2017}, our numerical algorithm conserves the total mass of the interacting bodies in a given collision, which is distributed between the largest remnant and the generated fragments. However, the model adopted by \citet{Mustill2018} incorporates a factor of mass removal assuming that most fragments are ground to smaller sizes in a collisional cascade, and then removed by radiation forces before they can be accreted on planetary embryos.

As for the switchover function $K$, \citet{Rein2019} investigated an inconsistency  found in \citet{Chambers1999} between the published function and the one that it is used in the available code. \citet{Rein2019} investigated different switching functions, in particular, two extreme cases: one infinitely differentiable function, and the Heaviside function. Their results show that an infinitely differentiable switching function does not perform much better over one close encounter than the polynomial function used by MERCURY. concluding that besides the discrepancy between the published switchover function and the , the MERCURY code is indeed a switching integrator. 

Finally, it is important to remark that the final water contents of the HZ planets produced in our runs are derived by post-processing N-body simulations. Future works should analyze the evolution of the water content of such planets from N-body experiments that track the removal and transfer of volatiles after each collision. Prescriptions derived from smoothed particle hydrodynamics experiments, such as those shown by \citet{Dvorak2015} and \citet{Burger2018}, should be included in future planetary formation N-body simulations in order to propose a more realistic model associated with the water evolution during the accretion stages.  

The incorporation of such considerations in future N-body experiments will allow us to obtain a better understanding concerning the physical properties of the terrestrial-like planets that composes the wide diversity of planetary systems of the Universe.

\begin{acknowledgements}
{We thank the anonymous referee for helpful comments improving this paper. We want to thank Dr. Alexander J. Mustill for useful comments about the numerical recipes implemented in our code. This work was partially financed by Consejo Nacional de Investigaciones Cient\'{\i}ficas y T\'ecnicas (CONICET) through PIP 0436/13, and Agencia de Promoci\'on Cient\'{\i}fica, through PICT 2014-1292, PICT 201-0505 and PICT 2016-2635. Moreover, we acknowledge the financial support by Facultad de Ciencias Astron\'omicas y Geof\'{\i}sicas de la Universidad Nacional de La Plata (FCAGLP-UNLP) and Instituto de Astrof\'{\i}sica de La Plata (IALP) for extensive use of their computing facilities.}
\end{acknowledgements}


\bibliographystyle{aa} 
\bibliography{Dugaro.bib} 

\end{document}